\documentclass[9pt,twocolumn,twoside]{osajnl}
\journal{osajnl}

\newcommand\numberthis{\addtocounter{equation}{1}\tag{\theequation}}

\makeatletter
\renewcommand*\env@matrix[1][\arraystretch]{%
  \edef\arraystretch{#1}%
  \hskip -\arraycolsep
  \let\@ifnextchar\new@ifnextchar
  \array{*\c@MaxMatrixCols c}}
\makeatother

\title{Hardware error correction for programmable photonics}

\author[1,*]{Saumil Bandyopadhyay}
\author[1,2]{Ryan Hamerly}
\author[1]{Dirk Englund}

\affil[1]{Research Laboratory of Electronics, MIT, 50 Vassar Street, Cambridge, MA 02139, USA}
\affil[2]{NTT Research Inc., PHI Laboratories, 940 Stewart Drive, Sunnyvale, CA 94085, USA}

\affil[*]{Corresponding author: saumilb@mit.edu}

\dates{}
\doi{}
\setboolean{displaycopyright}{false}

\begin{abstract}
\noindent Programmable photonic circuits of reconfigurable interferometers can be used to implement arbitrary operations on optical modes, facilitating a flexible platform for accelerating tasks in quantum simulation, signal processing, and artificial intelligence. A major obstacle to scaling up these systems is static fabrication error, where small component errors within each device accrue to produce significant errors within the circuit computation. Mitigating this error usually requires numerical optimization dependent on real-time feedback from the circuit, which can greatly limit the scalability of the hardware. Here we present a deterministic approach to correcting circuit errors by locally correcting hardware errors within individual optical gates. We apply our approach to simulations of large scale optical neural networks and infinite impulse response filters implemented in programmable photonics, finding that they remain resilient to  component error well beyond modern day process tolerances. Our results highlight a new avenue for scaling up programmable photonics to hundreds of modes within current day fabrication processes.
\end{abstract}

\begin{document}
\maketitle

\section{Introduction}

Integrated photonics is a key technology for optical communications and is advancing rapidly for applications in sensing, metrology, signal processing, and computation. Programmable photonic circuits of optical interferometers, which can implement arbitrary filters and passively compute matrix operations on optical modes, are the optical analogue to field programmable gate arrays (FPGAs) and enable photonic circuits to be flexibly reconfigured post-fabrication by software  \cite{bogaerts_programmable_2020, harris_linear_2018}. Experimental demonstrations of these circuits have already shown working systems operating on up to tens of optical modes, which have been used to accelerate tasks in quantum simulation \cite{harris_quantum_2017, wang_multidimensional_2018, qiang_large-scale_2018, sparrow_simulating_2018, carolan_universal_2015}, mode unscrambling \cite{miller_self-configuring_2013, miller_self-aligning_2013, annoni_unscrambling_2017, ribeiro_demonstration_2016, milanizadeh_manipulating_2019}, signal processing \cite{zhuang_programmable_2015, notaros_programmable_2017}, combinatorial optimization \cite{prabhu_accelerating_2020}, and artificial intelligence \cite{shen_deep_2017}.

While scaling up these systems to hundreds or thousands of modes would be immensely beneficial, doing so will require precise fabrication of tens of thousands of optical interferometers. Unfortunately, static component errors induced by process variation introduce errors that rapidly accrue for larger systems, limiting their usefulness for many applications. This is because the decomposition \cite{reck_experimental_1994, clements_optimal_2016} and optimization techniques used to program these circuits assume that all of the components are ideal; thus, any component errors result in a programming of the wrong operation. Component imprecision therefore has serious implications for the future of these systems; for example, beamsplitter variation as small as 2\%, which is a typical wafer-level variance \cite{mikkelsen_dimensional_2014}, has been shown to degrade accuracy by nearly 50\% for feedforward circuits used to implement classifiers for the MNIST image recognition task \cite{fang_design_2019}. Alternative programmable architectures, such as recirculating waveguide meshes consisting of triangular or hexagonal MZI lattices \cite{perez_field-programmable_2018, perez_reconfigurable_2016, zhuang_programmable_2015}, are similarly susceptible to component-induced error; device variation within these circuits introduces errors that will alter the response of phase-sensitive filters \cite{zand_effects_2020}. These systems' degree of sensitivity to component variation makes their control challenging when scaling up to large numbers of modes.

\begin{figure*}[tb]
    \centering
    \includegraphics[width=\textwidth]{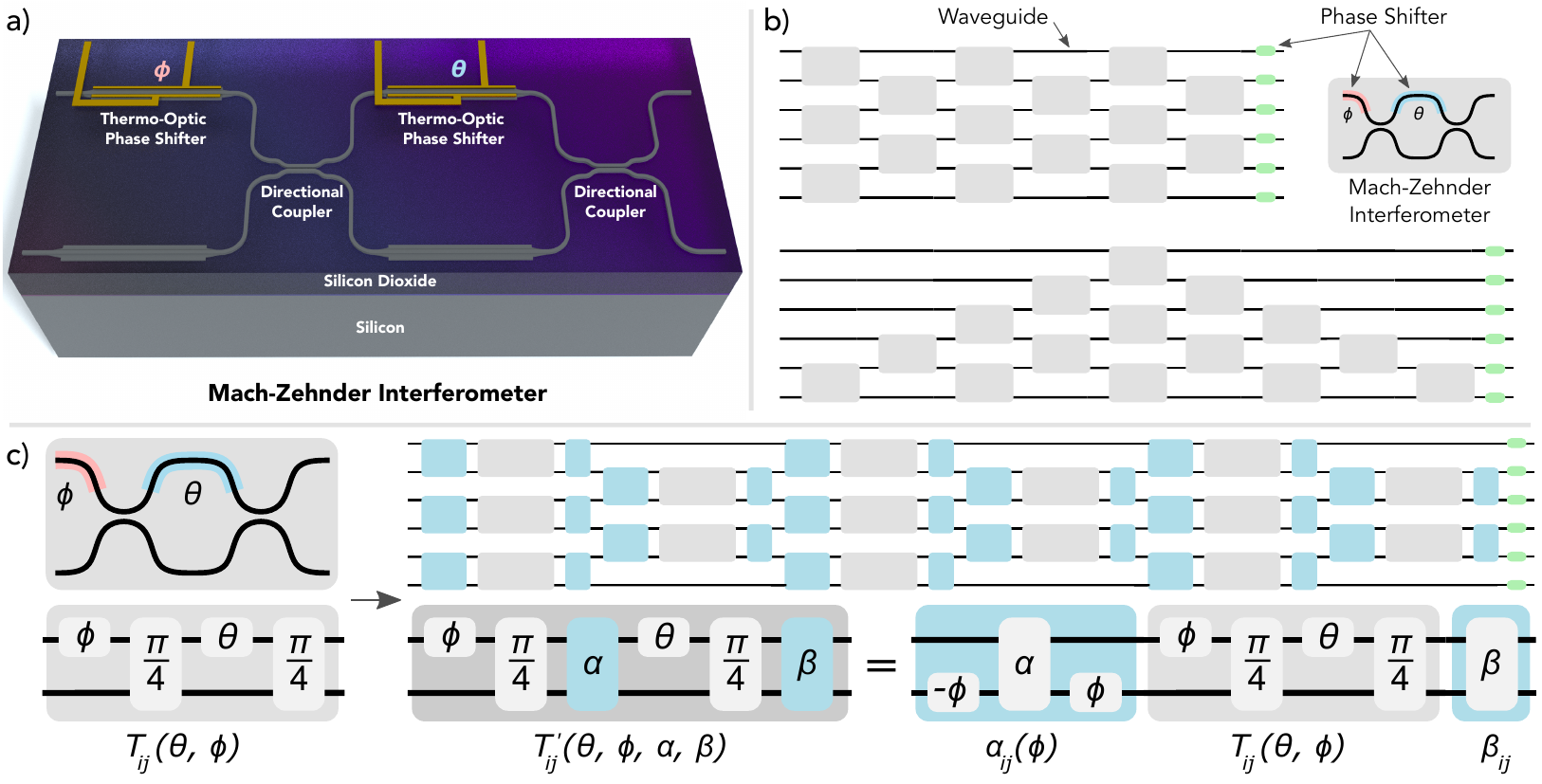}
    \caption{a) A Mach-Zehnder interferometer (MZI) on the silicon-on-insulator platform is composed of two 50-50 splitters, implemented with directional couplers, and an external phase shifter $\phi$ and internal phase shifter $\theta$ implemented with thermo-optic phase shifters. These devices act as electrically-controlled $2\times 2$ optical gates in programmable photonics. b) Arbitrary higher-dimensional matrix operations can be implemented by connecting $N(N-1)/2$ MZIs in a rectangular (top) or triangular (bottom) configuration. The Reck (triangular) and Clements (rectangular) decompositions \cite{reck_experimental_1994, clements_optimal_2016} describe a procedure for computing the phase settings for each MZI, but they assume the components are ideal. c) A realistic MZI implemented on a photonics platform will have splitting errors $\alpha, \beta$ for the two directional couplers within the interferometer. The effect of these hardware errors is to left- and right-multiply each programmable $2\times 2$ unitary $T_{ij}(\theta, \phi)$  implemented by an MZI by error matrices $\beta_{ij}, \alpha_{ij}(\phi)$. Applying the standard decomposition for ideal components to these imperfect optical gates will not produce the correct gate operation.}
    \label{overview}
\end{figure*}

Hardware errors are usually compensated for with numerical optimization. A number of global optimization approaches have been proposed in the past, including nonlinear optimization \cite{burgwal_using_2017, mower_high-fidelity_2015, lopez_auto-routing_2020, lopez_programmable_2020, perez-lopez_multipurpose_2020}, gradient descent \cite{pai_matrix_2019}, and \textit{in-situ} backpropagation and training for neural networks \cite{hughes_training_2018}. These strategies, however, are time-consuming and can scale poorly with circuit size. Moreover, it is often inefficient to retrain hardware settings for each individual chip. For many tasks, such as machine learning, model training is energy intensive; if the same model parameters are broadcast to thousands of chips within a data center, retraining the model for each  chip with a unique set of component imprecisions will be very costly. One can instead employ progressive algorithms making use of local feedback \cite{miller_setting_2017, pai_parallel_2020}; however, these algorithms, which iteratively optimize the settings of one device at a time, require $O(N^2)$ tap photodiodes to monitor the optical power within each individual interferometer. This requirement greatly increases the number of electrical lines and overall power consumption of the system.

This focus on \textit{in-situ} approaches reveals a critical roadblock for programmable photonics compared to electronic FPGAs. An FPGA does not optimize hardware settings in real time off readings taken directly from the chip; rather, control software takes it for granted that the logic gates are ideal and maps the requested function into a netlist that can be placed and routed within the chip. A similar capability for programmable photonics would greatly improve the scalability  of these systems; if this were the case, a desired optical function could be trained once on an idealized software model and ported over to many chips. The challenge for programmable photonics is that unlike FPGAs, photonic circuits are analog systems that are far more sensitive to errors within individual components. Enabling this level of scalability will therefore require the ability to deterministically correct hardware errors in photonic chips.

If a unitary operation is realizable by an imperfect photonic circuit, it should not require optimization to deduce the required settings; rather, a small perturbation in the device behavior due to component deviation should translate directly to a small perturbation in the interferometer's phase settings to recover the original unitary. This insight has led us to consider a local approach that corrects hardware errors one at a time within each optical gate composing the circuit. In this article, we present an approach to directly correct hardware errors for a programmable photonic circuit. Our algorithm outperforms previous approaches in several key respects: 1) it is flexible, requiring only a one time device calibration to directly compute the hardware settings for any given unitary; 2) for sufficiently low hardware errors the computed settings yield the exact unitary desired; and 3) our approach requires minimal overhead and does not make use of additional interferometers or internal detectors within every device. Our analysis is focused on feedforward programmable circuits that implement  arbitrary unitary matrices, as these systems have the most demanding requirements for fabrication precision. However, our approach is a local error correction strategy that individually corrects each $2\times2$ optical gate within the circuit. It therefore does not assume any particular structure to the circuit and can be generalized to any programmable architecture making use of interferometers, including feedforward circuits with redundant devices and recirculating waveguide meshes.

\section{Hardware Error Correction}

Local error correction requires characterization of each phase shifter and passive splitter in the photonic circuit. The calibration is performed once with the results stored in a lookup table; any arbitrary function can then be programmed by computing the settings for an ideal set of MZIs and converting them, one by one, to the corresponding settings for an imperfect device. In the Supplementary Information (Sec. I.), we describe how to calibrate these parameters using detectors only at the circuit outputs; assuming these errors are known, we can proceed with error correction as follows.

The fundamental optical gate of a programmable photonic circuit is a $2 \times 2$ Mach-Zehnder interferometer (MZI) composed of an external phase shifter on one input, two 50-50 beamsplitters, and an internal phase shifter on one of the modes between the  splitters (Fig.~\ref{overview}a). This device is an electrically programmable beamsplitter capable of performing a $2 \times 2$ unitary operation $T_{ij}(\theta, \phi)$ on optical modes $i,j$ parameterized by the external phase shift $\phi$ and the internal phase shift $\theta$. 

On an integrated photonics platform, the 50-50 splitters can be realized by a directional coupler or multimode interferometer (MMI); the operation of these splitters can be described by a $2 \times 2$ matrix:
\begin{equation}
\begin{bmatrix} \cos (\pi/4 + \alpha) & i\sin(\pi/4 + \alpha) \\ i \sin (\pi/4 + \alpha) & \cos (\pi/4 + \alpha) \end{bmatrix}
\end{equation}
where $\alpha$ describes the deviation from an ideal 50-50 splitting behavior. For an ideal splitter $\alpha=0$, this matrix reduces to:
\begin{equation}
    \frac{1}{\sqrt{2}} \begin{bmatrix} 1 & i \\ i & 1 \end{bmatrix}
\end{equation}
The overall operation $T_{ij}(\theta, \phi)$ performed by a single ideal MZI is therefore:
\begin{align*}
    T_{ij}(\theta, \phi) &=
    \frac{1}{2}
    \begin{bmatrix} 1 & i \\ i & 1 \end{bmatrix}
    \begin{bmatrix} e^{i\theta} & 0 \\ 0 & 1 \end{bmatrix}
    \begin{bmatrix} 1 & i \\ i & 1 \end{bmatrix}
    \begin{bmatrix} e^{i\phi} & 0 \\ 0 & 1 \end{bmatrix} \\
    &= i e^{i \theta/2} \begin{bmatrix} e^{i \phi} \sin (\theta/2) & \cos (\theta/2) \\ e^{i \phi} \cos (\theta/2) & - \sin (\theta/2) \end{bmatrix} \numberthis
\end{align*}
where $\theta, \phi$ are single-mode phase shifts on the top arm. 

Higher dimensional matrix operations can be implemented with this unit cell by applying the Clements \cite{clements_optimal_2016}  and Reck \cite{reck_experimental_1994}  decompositions (Fig.~\ref{overview}b). These algorithms decompose an arbitrary $N$-dimensional unitary $U$ into a product of $N(N-1)/2$ two-dimensional unitaries computed by interference between nearest-neighbor optical modes, followed by phase shifts on the output modes corresponding to a diagonal matrix $D$:
\begin{equation}
    U = D \prod T_{ij} (\theta, \phi)
\end{equation}

We now analyze the impact of fabrication error. If the MZI has imperfect splitters with errors $\alpha, \beta$, the operation of the MZI must now be parameterized with four variables $T^\prime_{ij}(\theta, \phi, \alpha, \beta)$ (Fig.~\ref{overview}c):
\begin{align}
     i e^{i \theta / 2} &\begin{bmatrix}[2.5]
        \begin{matrix}[1]e^{i \phi} (\cos(\alpha - \beta) \sin (\theta/2)~+\\ i  \sin (\alpha + \beta) \cos (\theta/2)) \end{matrix}  & \begin{matrix}[1] \cos(\alpha + \beta) \cos (\theta/2)   ~+\\ i \sin (\alpha - \beta) \sin (\theta/2) \end{matrix} \\
        \begin{matrix}[1] e^{i \phi} (\cos(\alpha + \beta) \cos (\theta/2)~-\\i \sin (\alpha - \beta) \sin (\theta/2) ) \end{matrix} &  \begin{matrix}[1]- \cos(\alpha - \beta) \sin (\theta/2)~+\\i \sin (\alpha + \beta) \cos (\theta/2) \end{matrix}
     \end{bmatrix} \\
        = &\begin{bmatrix} \cos \beta & i \sin \beta \\ i \sin \beta & \cos \beta \end{bmatrix}
         \hat{T}(\theta, \phi)
         \begin{bmatrix} \cos \alpha & i e^{-i \phi} \sin \alpha \\ ie^{i \phi} \sin \alpha & \cos \alpha \end{bmatrix} 
\end{align}
In the limit $\alpha, \beta \rightarrow 0$, the second term of each entry in the matrix $T^\prime_{ij}(\theta, \phi, \alpha, \beta)$ drops out and we recover the expected transformation for an ideal device. Naturally, implementing the usual decomposition on these imperfect devices will not yield the desired unitary:
\begin{align}
    D \prod T^\prime_{ij} (\theta, \phi, \alpha, \beta) \neq D \prod T_{ij} (\theta, \phi)
\end{align}
To program into an imperfect circuit a desired unitary $U = \prod T_{ij} (\theta, \phi)$, we apply local corrections $\theta \rightarrow \theta^\prime, \phi \rightarrow \phi^\prime$ to each device such that $T^\prime_{ij} (\theta^\prime, \phi^\prime, \alpha, \beta) = T_{ij} (\theta, \phi)$. 

\begin{figure*}[tb]
    \centering
    \includegraphics[width=\textwidth]{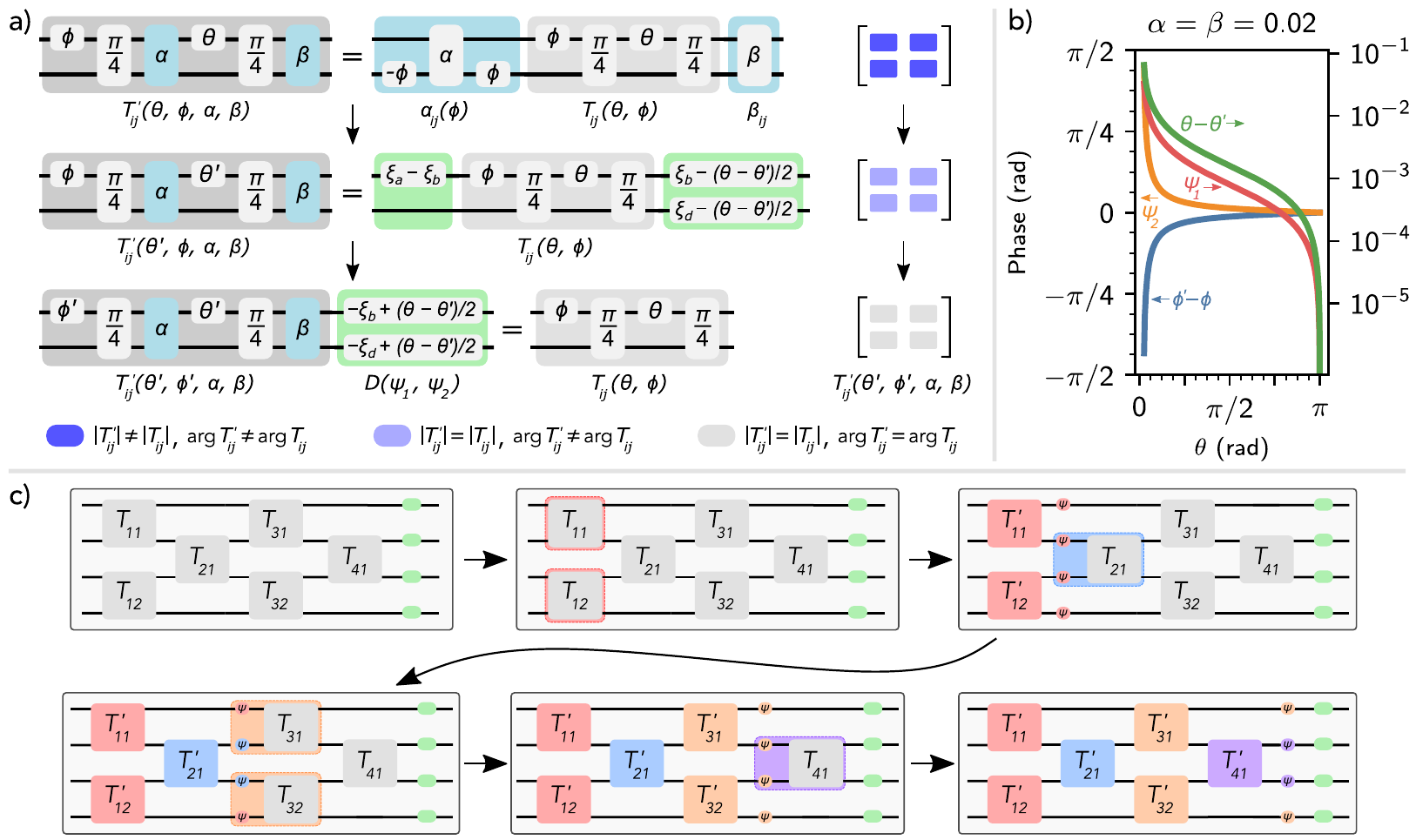}
    \caption{a) Fabrication-induced errors within each MZI  can be corrected by applying local corrections $\theta \rightarrow \theta^\prime$, $\phi \rightarrow \phi^\prime$ to the device. We first correct $\theta$ to set the magnitudes of the elements of $T_{ij}$ equal to $T_{ij}^\prime$. Once the amplitude terms are set correctly, we apply phase corrections to the input and outputs of the device to correct phase errors between $T_{ij}$ and $T_{ij}^\prime.$ b) The corrections $\phi^\prime - \phi$, $\theta- \theta^\prime$, $\psi_1$, $\psi_2$ applied to an MZI with two 52-48 beamsplitters ($\alpha=\beta=0.02$). The arrows on the plot indicate which vertical axis each curve corresponds to. c) The procedure for programming a unitary with hardware errors on a $4\times 4$ rectangular unitary circuit. We first program each MZI to the $(\theta, \phi)$ setting obtained with the standard decomposition in \cite{clements_optimal_2016}. Each MZI is then converted $T_{ij} \rightarrow T_{ij}^\prime$ to the settings for an imperfect device one column at a time. At each step we propagate the output phase shifts $\psi_1, \psi_2$ forward in the circuit until the entire network is corrected.}
    \label{ec}
\end{figure*}

Figure \ref{ec}a illustrates our approach. We begin by finding $\theta^\prime$ such that the magnitudes of the entries of $T^\prime_{ij} (\theta^\prime, \phi^\prime, \alpha, \beta)$ equal those of $T_{ij} (\theta, \phi)$. This condition produces the following expression for $\theta^\prime$ (Supplementary Info., Sec. III.A.):
\begin{align}
    \theta^\prime &= 2 \arcsin \sqrt{\frac{\sin^2 (\theta/2) - \sin^2 (\alpha + \beta)}{\cos^2(\alpha - \beta) - \sin^2(\alpha+\beta)}} \numberthis
    \label{theta_correct}
\end{align}
Component errors restrict the range over which $\theta$ is physically realizable. The above expression has a solution only if $\sin^2 (\theta/2) > \sin^2 (\alpha + \beta)$ and if $\sin^2 (\theta/2) < \cos^2 (\alpha - \beta)$. This restricts $\theta$ to the range:
\begin{equation}
    2|\alpha + \beta| < \theta < \pi - 2|\alpha-\beta| \label{theta_range}
\end{equation}
If the matrix decomposition requires $\theta$ outside this range, the error is minimized by setting $\theta^\prime=0$ (if $\theta < 2\lvert \alpha+\beta \rvert$) or $\theta^\prime=\pi$ (if $\theta > \pi - 2 \lvert \alpha - \beta \rvert$). 

Assuming we can physically implement the required value of $\theta^\prime$, the magnitudes of the elements of $T^\prime_{ij} (\theta^\prime, \phi^\prime, \alpha, \beta)$ and $T_{ij} (\theta, \phi)$ are now the same, but each element of $T^\prime_{ij}$ will have an undesired extraneous phase $\xi_a,\xi_b,\xi_c,\xi_d$ relative to the corresponding term in $T_{ij}$ that must be corrected. We can therefore rewrite $T^\prime_{ij} (\theta^\prime, \phi^\prime, \alpha, \beta)$ as
\begin{align}
    T^\prime_{ij} &= 
    i e^{i \theta^\prime/2} \begin{bmatrix} e^{i \phi^\prime} e^{i \xi_a} \sin (\theta/2)  & e^{i \xi_b} \cos (\theta/2)  \\ e^{i \phi^\prime} e^{i \xi_c} \cos (\theta/2)  & - e^{i \xi_d} \sin (\theta/2)  \end{bmatrix} \\
    &= i e^{i \theta^\prime/2}
    \begin{bmatrix} e^{i \xi_b} & 0 \\ 0 & e^{i \xi_d} \end{bmatrix} 
    \begin{bmatrix} e^{i (\phi^\prime + \xi_a - \xi_b)} \sin (\theta/2)  &  \cos (\theta/2)  \\ e^{i (\phi^\prime + \xi_a - \xi_b)} \cos (\theta/2)  & - \sin (\theta/2)  \end{bmatrix}
\end{align}
where the simplification in the second line originates from unitarity requiring that $\xi_a + \xi_d = \xi_b + \xi_c$. We correct the phase errors in $T^\prime_{ij}$ by setting $\phi^\prime = \phi + \xi_b - \xi_a$ and by applying additional phases $\psi_1 = -\xi_b + (\theta - \theta^\prime)/2$, $\psi_2 = -\xi_d + (\theta - \theta^\prime)/2$ to the top and bottom output modes, respectively. Applying these corrections will set $T^\prime_{ij} (\theta^\prime, \phi^\prime, \alpha, \beta)$ exactly equal to $T_{ij} (\theta, \phi)$.

Expressions for the phase errors $\xi_a, \xi_b, \xi_d$ can be constructed by setting the complex arguments of the elements of $T_{ij}$ equal to those of $T^\prime_{ij}(\theta^\prime, \phi^\prime, \alpha, \beta)$. From this, we find that:
\begin{align*}
    \phi^\prime = \phi &+ \arctan \left [ \frac{\sin(\alpha-\beta)}{\cos(\alpha+\beta)} \tan (\theta^\prime/2) \right ]\\ &- \arctan \left [ \frac{\sin(\alpha+\beta)}{\cos(\alpha-\beta)} \cot (\theta^\prime/2) \right ]
    \numberthis
    \label{phi_prime}
\end{align*}
\begin{equation}
    \psi_1 = -\arctan \left [ \frac{\sin(\alpha-\beta)}{\cos(\alpha+\beta)} \tan (\theta^\prime/2) \right ] + (\theta - \theta^\prime)/2 \numberthis
    \label{psi_1}
\end{equation}
\begin{equation}
    \psi_2 = \arctan \left [ \frac{\sin(\alpha+\beta)}{\cos(\alpha-\beta)} \cot (\theta^\prime/2) \right ] + (\theta - \theta^\prime)/2 \numberthis
    \label{psi_2}
\end{equation}

The errors $\theta-\theta^\prime, \phi^\prime-\phi, \psi_1, \psi_2$ as a function of $\theta$ for an example MZI with two 52-48 ($\alpha=\beta=0.02$) splitters are shown in Figure \ref{ec}b. While the corrections to $\theta$ and $\psi_1$ are small ($\sim 0.1$ rad), the errors for $\phi$ and $\psi_2$ are quite substantial. In particular, for low device reflectivities ($\theta \approx 0$), the phase corrections required can exceed 1 rad.

Generally, we cannot apply the auxiliary phases $\psi_1, \psi_2$ locally to the device being corrected, since the output modes do not have phase shifters. In most cases, one of the two can be incorporated into the external phase shifter setting of an MZI in the subsequent column. The other phase can be applied by observing that:
\begin{equation}
    T_{ij}(\theta, \phi) \begin{bmatrix} e^{i \psi_1} & 0 \\ 0 & e^{i \psi_2} \end{bmatrix} = \begin{bmatrix} e^{i \psi_2} & 0 \\ 0 & e^{i \psi_2} \end{bmatrix} T_{ij}(\theta, \phi + \psi_1 - \psi_2)
    \numberthis \label{forward_propagate}
\end{equation}
Using this fact, we can propagate the auxiliary phases forward, through all of the columns of the network, out to the phase shifters $D$ located on the output modes of the circuit. This procedure, illustrated in Figure \ref{ec}c, produces a modified output phase screen $D^\prime$ such that:
\begin{equation}
    U = D \prod T_{ij} (\theta, \phi) = D^\prime \prod T^\prime_{ij} (\theta^\prime, \phi^\prime, \alpha, \beta)
\end{equation}
 
 Depending on the component imperfections and the required value of $\theta$, we may also be able to program $\theta^\prime$ such that $|T^\prime_{ij} (\theta^\prime, \phi^\prime, \alpha, \beta)| = |T_{ij} (\theta, \phi)|$ if the condition in equation (\ref{theta_range}) is satisfied. If every MZI in the circuit satisfies the condition in equation (\ref{theta_range}), we can recover the exact unitary desired. However, if some MZIs in the circuit cannot realize the required splitting, that exact unitary is not physically realizable by the device. In this case, correcting the phases $\phi^\prime, \psi_1, \psi_2$ and setting $\theta^\prime$ as close to the required value as possible minimizes the gate error $\lVert T_{ij} - T^\prime_{ij} \rVert$.

We can summarize the algorithm for programming of a matrix $U$ as follows:
\begin{enumerate}
    \item Calculate the required values for $\theta, \phi$ assuming ideal components, using the procedure described by Reck \cite{reck_experimental_1994} or Clements \cite{clements_optimal_2016}.
    \item For each device, set $\theta \rightarrow \theta^\prime$ using the expression in equation (\ref{theta_correct}). If $\theta < 2|\alpha + \beta|$, set $\theta^\prime = 0$; if $\theta > \pi - 2 \lvert \alpha-\beta \rvert$, set $\theta^\prime = \pi$.
    \item Apply phase corrections $\phi^\prime, \psi_1, \psi_2$ as given in equations \ref{phi_prime}-\ref{psi_2}. Propagate $\psi_1, \psi_2$ forward to the output phase screen $D$ with the expression in equation (\ref{forward_propagate}).
\end{enumerate}

We have illustrated this procedure for the example of feedforward unitary circuits, but the same principles apply for other  architectures. Each optical gate within any programmable circuit can be corrected to the required $2 \times 2$ unitary operation $T_{ij}$ with the aforementioned procedure. The expressions provided assume a specific form for the MZI (Fig.~\ref{overview}), but they can be easily modified to apply to other designs, such as the dual-drive tunable basic unit (TBU) used in recirculating architectures \cite{perez-lopez_integrated_2019}.

\section{Discussion}

\subsection{Hardware Performance}

We analyzed the performance of error correction through numerical simulations of programmable photonic circuits with fabrication imperfections. Results were obtained with a custom simulation package written using \verb|NumPy| \cite{harris_array_2020}. Further details are included in the Supplementary Information.

\begin{figure}[tb]
    \centering
    \includegraphics[width=\columnwidth]{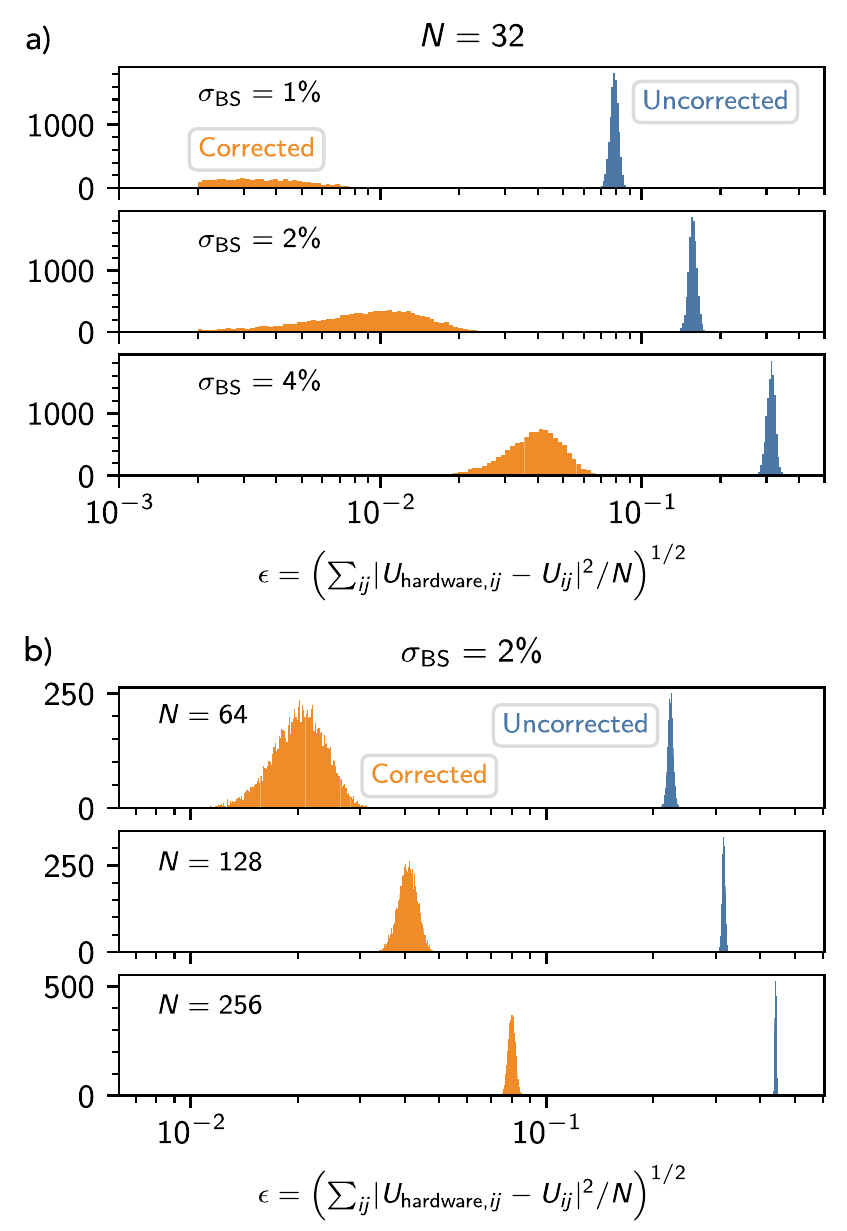}
    \caption{a) Matrix error $\epsilon$ before and after correction for 100 random unitaries implemented on 100 random circuits with varying beamsplitter statistics. b) Matrix error $\epsilon$ before and after correction for $N=\{64, 128, 256\}$ with a beamsplitter variation $\sigma_{BS}=2\%$.}
    \label{matrix_error}
\end{figure} 

Figure \ref{matrix_error}a shows the matrix error (relative error per entry) $\epsilon = (\sum_{ij} \lvert U_{\text{hardware}, ij} - U_{ij} \rvert^2/N)^{1/2}$ for 100 Haar random unitaries implemented on 100 randomly generated $N=32$-mode unitary circuits with mean beamsplitter transmission $\eta = (50 \pm \sigma_{\text{BS}})\%$. The beamsplitter errors are  independently sampled from a Gaussian distribution; for large $N$, the distribution shape will not greatly affect the results. We find that error correction reduces $\epsilon$ significantly, sometimes by more than an order of magnitude. This improvement is larger for circuits with small splitting errors, as they are more likely to satisfy equation (\ref{theta_range}) and program the required $\theta^\prime$ for all devices within the circuit. However, even for circuits with large $\sigma_\text{BS}$, where many MZIs may not be programmable to the required $\theta$, the improvement in $\epsilon$ is substantial as all  errors in $\phi, \psi_1, \psi_2$ can always be corrected.

In Figure \ref{matrix_error}b, we show $\epsilon$ with and without error correction for circuit sizes $N=\{64, 128, 256\}$. For these simulations, we chose a beamsplitter variation of $\sigma_{BS} = 2\%$, which is a typical wafer-level variance \cite{mikkelsen_dimensional_2014}. While the improvement in $\epsilon$ diminishes for larger $N$, we still find substantial improvement gained in our approach for up to 256 modes. For large unitary circuits most MZIs need to be programmed to reflectivities close to $\theta \approx 0$ \cite{russell_direct_2017}; the increasing fraction of devices that cannot be programmed to the required splitting accounts for the increase in $\epsilon$ with $N$. Nevertheless, there is always some improvement in $\epsilon$, as any phase errors introduced by the components can be corrected. Our results suggest that substantial performance improvements can still be achieved by error correction for circuits with hundreds of modes, which is well beyond the size of the current state-of-the-art ($N=64$) in programmable photonics \cite{harris_accelerating_2020}.

\subsection{Application: Optical Neural Networks on Feedforward Programmable Circuits}

\begin{figure}[tb]
    \centering
    \includegraphics[width=\columnwidth]{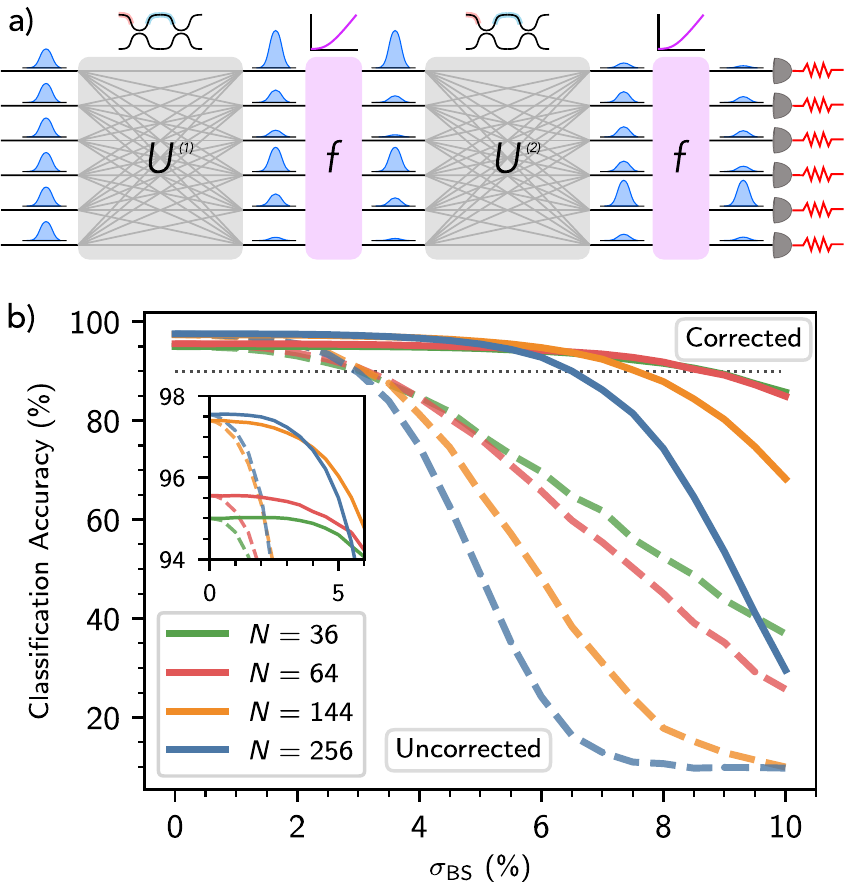}
    \caption{a) We simulated the effect of component error on two-layer optical neural networks for the MNIST task. Matrix-vector products are calculated optically in the photonic circuit, and modReLU-like activation functions are implemented electro-optically \cite{arjovsky_unitary_2016, williamson_reprogrammable_2020}. b) Median accuracy for 300 unitary circuits as a function of $\sigma_{BS}$ with and without correction for a photonic image classifier for the MNIST task with $N=\{36, 64, 144, 256\}$ neurons. Error correction significantly improves the fabrication tolerance of the neural network to beyond current-day process tolerances, even for systems with hundreds of modes. As the inset shows, even circuits with 4\% splitter error preserve the baseline performance within 1\%.}
    \label{onn_data}
\end{figure} 

To further benchmark the performance of our error correction protocol, we applied this approach to simulations of a programmable photonic system, namely a two-layer neural network conducting inference with a feedforward programmable photonic circuit. The architecture of the neural network is similar to that studied in \cite{shen_deep_2017, pai_parallel_2020, fang_design_2019}, where forward inference is optically computed through passive interference within a unitary photonic circuit coupled with an electrical or electro-optic nonlinearity \cite{williamson_reprogrammable_2020}. Optical machine learning is a key application area for photonic error correction, as model training is both time-consuming and energy intensive, making it impractical to retrain on each individual piece of hardware with a unique set of fabrication errors. Preferably, a model would be highly optimized once in software, after which corrections are applied within the hardware to restore the original software-trained model from any fabrication-induced errors.

The neural networks we benchmark are based on the architecture described in \cite{pai_parallel_2020}. Using the \verb|Neurophox| package, we trained two-layer neural networks with $N=\{36, 64, 144, 256\}$ neurons to recognize low-frequency Fourier features of handwritten digits from the MNIST task. The activation function between layers was assumed to be a modReLU function implemented using an electro-optic nonlinearity \cite{arjovsky_unitary_2016, williamson_reprogrammable_2020}. Further details on the  network architecture and training are included in the Supplementary Information.

Figure \ref{onn_data} shows the median classification accuracy for 300 randomly generated circuits as a function of the beamsplitter statistics  $\eta = (50 \pm \sigma_{\text{BS}})\%$. The smaller circuits ($N=36, 64$) exhibit roughly $95-96\%$ accuracy after training, while the larger circuits ($N=144, 256$) exhibit a slightly higher model accuracy of  $\sim 97\%$. The larger circuits, however, are less resilient to errors;  without error correction classification accuracy drops to below  90\% for all circuit sizes at a splitter variation as low as $\sim 3\%$.

Hardware error correction extends this cutoff to more than $6\%$, which is well beyond modern-day process tolerances \cite{mikkelsen_dimensional_2014}. Moreover, without correction classification accuracy drops significantly at even typical wafer-level variances (2\%). However, with error correction there is almost no drop in accuracy at these variances and less than 1\% accuracy loss for beamsplitter variations as high as 4\%. We expect this margin for fabrication error will prove important as optical neural networks scale up. These results suggest that error correction in programmable photonics can enable high-accuracy neural networks of up to hundreds of modes within current-day process tolerances. 

\subsection{Application: Tunable Dispersion Compensators on Recirculating Waveguide Meshes}

\begin{figure}[tb]
    \centering
    \includegraphics[width=\columnwidth]{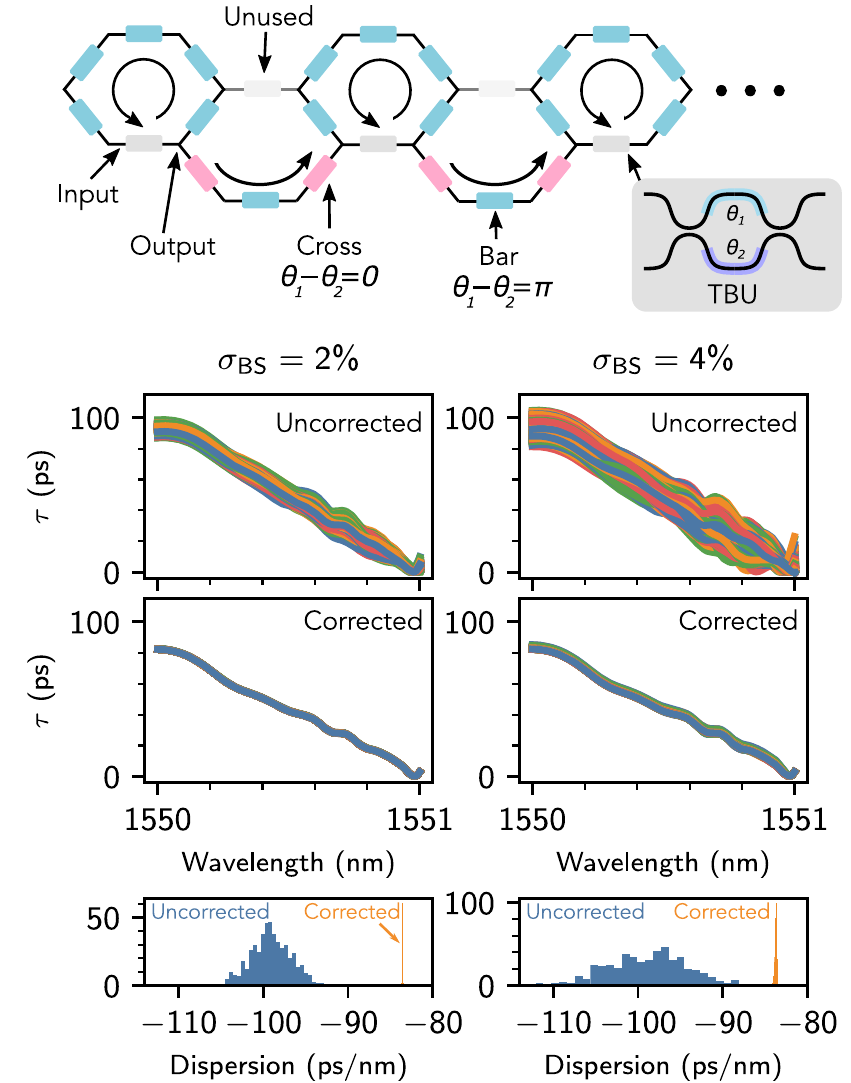}
    \caption{Top: Simulations of a tunable dispersion compensator (TDC) implemented on a recirculating waveguide mesh with 15 tunable-coupling ring resonators coupled serially to one another. Bottom: After training the mesh parameters to implement a fixed linear group delay dispersion on an ideal model, small beamsplitter errors will introduce variations in the implemented group delay $\tau$ profile. Plotted are the group delay profiles for 500 randomly generated circuits before and after correction. Correcting the settings of each TBU restores the desired performance, eliminating the need to retrain on the hardware. Also displayed is the distribution of the group delay dispersion before and after correction.}
    \label{tdc}
\end{figure} 

While our analysis has focused on feedforward programmable photonic meshes, our results can also be applied to recirculating architectures useful in RF and optical signal processing. These recirculating meshes, which are usually configured in hexagonal or triangular lattices, enable implementation of finite impulse response (FIR) and infinite impulse response (IIR) filters by configuring waveguides into asymmetric MZIs and ring resonators, respectively \cite{zhuang_programmable_2015, perez_reconfigurable_2016, perez_field-programmable_2018}. Unlike the feedforward architectures, the programming of these structures usually cannot be determined analytically and must be found through optimization \cite{lopez_auto-routing_2020, lopez_programmable_2020, perez-lopez_multipurpose_2020}. Since optimization can be time-consuming for complex systems, error correction can enable optimizing these circuit parameters on idealized models and then porting them over to hardware without retraining. 
As an example, we simulated the performance of an IIR filter functioning as a tunable dispersion compensator (TDC) on a hexagonal waveguide lattice \cite{perez_reconfigurable_2016}. TDC modules are of interest for numerous applications, including compensating chromatic dispersion in optical communication links \cite{madsen_optical_1998} and enabling high-dimensional quantum key distribution (QKD) with temporal modes \cite{mower_high-dimensional_2013}.

We implemented the TDC using an architecture similar to the tunable-coupling ring array described in \cite{notaros_programmable_2017}. Programmable dispersion is achieved by individually tuning the coupling and resonance of each ring in a chain of 15 resonators coupled serially to one another. Each ring is implemented with a single MZI (often referred to as the tunable basic unit, or TBU) in a hexagonal mesh acting as the coupler, while five other TBUs are programmed to the bar state to implement feedback. For simplicity we do not simulate routing within the hexagonal mesh, but instead simulate the transfer function of each individual filter implemented using TBUs with fabrication imperfections. Using the constrained optimization by linear approximations (COBYLA) routine in \verb|SciPy| \cite{powell_direct_1998, scipy_10_contributors_scipy_2020}, we trained the TBU parameters on an idealized model to implement a group delay dispersion of $-85$ ps/nm over the bandwidth of a 50 GHz ITU channel.

Figure \ref{tdc} shows the group delay $\tau$ profiles for 500 randomly generated TDC modules implemented using TBUs with $\sigma_{\text{BS}} = \{2, 4\}\%$ before (top) and after (bottom) error correction. Similar to optical neural networks, precise implementation of a TDC requires accurate phase control throughout the circuit. Fabrication errors introduce spurious phases at each resonance, which results in significant variation of the dispersion profile for even slight component errors. As our results show, correcting the parameters of each TBU locally is sufficient to restore the desired dispersion profile. 

While we can correct the coupling and phase parameters for each ring, errors $\alpha, \beta$ also introduce some loss at each TBU programmed to the bar state, as the bar transmission is reduced to $\cos^2(\alpha-\beta)$. The remainder of the light is directed into unused couplers in the circuit, effectively incurring loss. This alters the critical coupling condition, which results in the slight variation in the dispersion profile observed for $\sigma_{BS} = 4\%$. Our simulations assume $\alpha, \beta$ are independent, Gaussian random variables; in practice, however, $\alpha, \beta$ for a single device are strongly correlated \cite{yang_phase_2015, lu_performance_2017}. Therefore, our simulations likely overestimate the loss incurred at each TBU programmed to the bar state.

\subsection{Scalability and Outlook}
We have presented an approach for characterizing and correcting for hardware errors in programmable photonic circuits. To conclude, we analyze the expected improvement our technique enables and how it will perform as these circuits scale up.

For a unitary photonic circuit, applying the Reck or Clements decompositions produces an average matrix error $\epsilon$ of (Supplementary Info., Sec. II.A.):
\begin{equation}
    \langle \epsilon \rangle \approx \sigma_{BS} \sqrt{2(N-1)} 
\end{equation}
If we can correct all errors in $\theta$, then $\epsilon_\text{corrected} \rightarrow 0$. We can therefore estimate the expected $\epsilon_\text{corrected}$ by computing the fraction of MZIs that cannot be programmed to the required splitting value, i.e. the condition in equation (\ref{theta_range}).

The distribution of phase shifter settings for a unitary circuit can be related to the Haar measure on the unitary group \cite{russell_direct_2017}. The probability an MZI is programmed to a value $\theta < \xi$ is (Supplementary Info., Sec. II.C.):
\begin{align}
    P(\theta < \xi)&= \sum_{k=1}^{N-1} \frac{2(N-k)}{N(N-1)} (1-\cos^{2k}(\xi/2))\\
    &\approx \frac{N+1}{12}\xi^2
\end{align}
We disregard the probability an MZI is programmed to a splitting  $\theta > \pi - 2|\alpha-\beta|$, which is negligibly small for large $N$ \cite{russell_direct_2017}. Error correction cannot fix the splitting error if  $\theta < 2\lvert \alpha + \beta \rvert$; therefore:
\begin{align}
    \langle \epsilon_\text{corrected} \rangle &\approx \left (\frac{1}{2} P(\theta < 2\lvert \alpha + \beta \rvert) \langle \epsilon^2 \rangle \right)^{1/2} \\
    &=  \sigma_\text{BS}^2 \sqrt{\frac{2(N^2-1)}{3}} 
\end{align}
We find that error correction effectively reduces the hardware error from $\epsilon$ to $\approx (1/\sqrt{6})\epsilon^2$. The expected error improvement is:
\begin{equation}
    \frac{\langle \epsilon \rangle }{\langle \epsilon_\text{corrected} \rangle} \approx \frac{\sqrt{3}}{\sigma_\text{BS}\sqrt{N+1}}
\end{equation}

\begin{figure}[tb]
    \centering
    \includegraphics[width=\columnwidth]{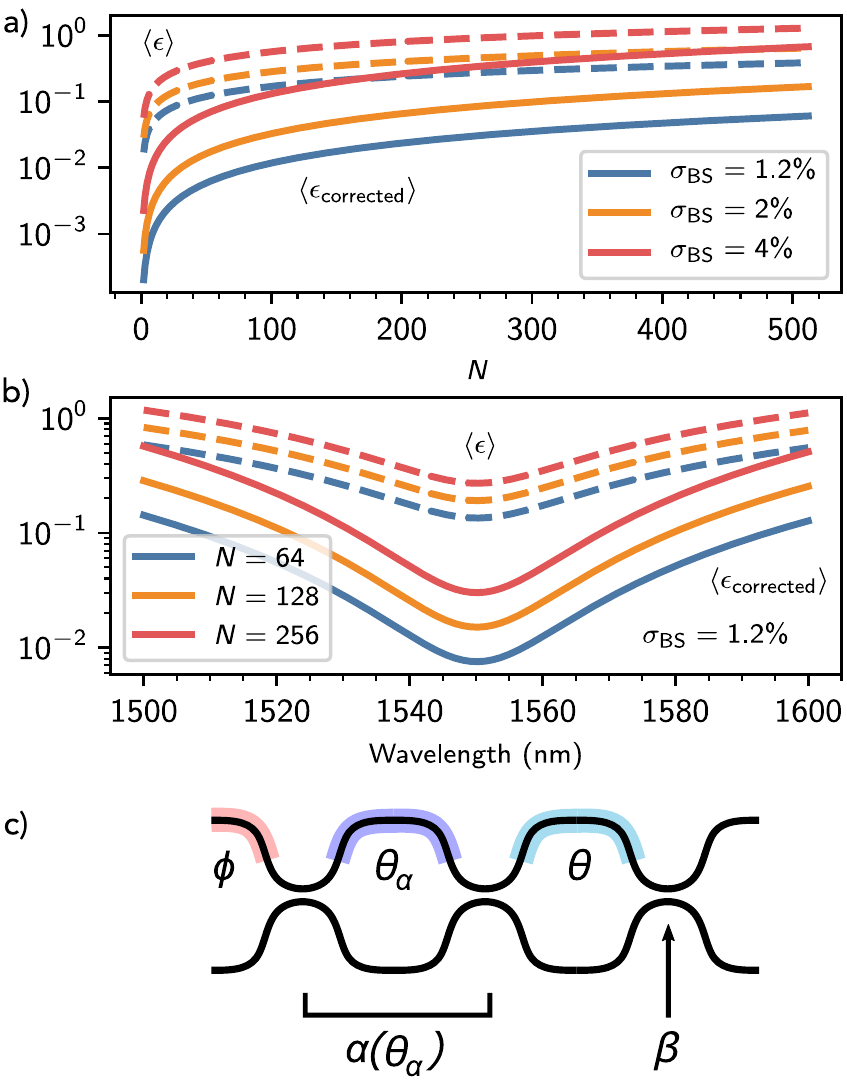}
    \caption{a) $\langle \epsilon \rangle, \langle \epsilon_\text{corrected} \rangle$ as a function of circuit size $N$ for $\sigma_\text{BS} = \{1.2, 2, 4\}\%$. b) Average circuit error as a function of wavelength for $N=\{64,128,256\}$ using the optimal directional coupler design in \cite{mikkelsen_dimensional_2014}. c) Redundant MZI for implementing perfect optical gates. One of the two beamsplitters is an MZI that can be tuned to implement an error $\alpha(\theta_\alpha)$ which  compensates for the error $\beta$.}
    \label{scalability}
\end{figure} 

$\langle \epsilon \rangle$ and $\langle \epsilon_\text{corrected} \rangle$ as a function of $N$ are plotted in Figure \ref{scalability}a. We consider $\sigma_\text{BS} = 1.2\%$, which is the state-of-the-art reported in \cite{mikkelsen_dimensional_2014}, as well as more relaxed tolerances $\sigma_\text{BS} = \{2, 4\}\%$. For $\sigma_\text{BS}$ as high as 4\%, error correction produces at least a factor of two (and often more) improvement in the error for circuits as large as $N=500$. We therefore expect our approach to have wide applicability in the near term as the size of programmable photonic circuits scale up.    

Error correction also greatly improves the optical bandwidth of unitary circuits. Since directional couplers are highly wavelength sensitive, dense wavelength-division multiplexing (DWDM) requires re-fabricating the same circuit with components optimized at each wavelength channel. Our approach, however, enables the use of the same hardware across a wide wavelength range. In Figure~\ref{scalability}b we show the expected hardware errors for large circuits across a 100 nm bandwidth using the optimal splitter ($\sigma_\text{BS} = 1.2\%$) design in \cite{mikkelsen_dimensional_2014}. We find that the corrected error for an $N=256$ circuit across a 60 nm bandwidth (1520-1580 nm) will be lower than the \textit{uncorrected} error at the design wavelength $\lambda = 1550$ nm. Even lower errors could be achieved using multimode interference (MMI) couplers; these devices have large bandwidths but often suffer from static splitting imbalances \cite{guan_compact_2017}, i.e., $\alpha, \beta$ are invariant to wavelength, but $\langle \alpha \rangle, \langle \beta \rangle \neq 0$. A circuit with large-bandwidth MMI couplers can thus use error correction to achieve a large instantaneous bandwidth, for instance to compute over many parallel wavelength channels.

The results in Figure \ref{scalability}a suggest a fundamental error bound achievable with local correction for unitary circuits. Our approach yields comparable results to those achieved with self-configuration procedures \cite{miller_self-aligning_2013, pai_parallel_2020} but does not require a specific structure for the circuit or photodiodes within each device. If the condition in equation (\ref{theta_range}) is satisfied, local correction obtains $\epsilon_\text{corrected} = 0$ in $O(1)$ time. If this condition is not satisfied, it is sometimes possible to achieve a larger reduction in error with a \textit{global} optimization approach \cite{burgwal_using_2017, pai_matrix_2019}. However, these approaches, which require photodiodes within each device or output measurements whose number scale nonlinearly with the number of modes, become increasingly inaccessible experimentally as \textit{N} scales up. Local correction requires minimal overhead and can guarantee a minimum error given certain guarantees on the component performance, making it ideal for standardizing  performance across large numbers of chips.

Moreover, this error bound applies only to feedforward, unitary circuits with no redundant devices. $\epsilon$ lower than this bound can be achieved by incorporating additional, redundant MZIs; for instance, one can implement ``perfect'' optical gates by incorporating an additional phase shifter into the MZI, as shown in Figure \ref{scalability}c. This device can be trained with optimization to implement any desired unitary $T_{ij}(\theta, \phi)$ perfectly \cite{suzuki_ultra-high-extinction-ratio_2015, wang_tolerant_2020}. The error correction formalism enables calculation of these settings analytically. One of the two constituent splitters is a passive component with error $\beta$, while the other splitter is an MZI that implements a tunable error $\alpha(\theta_\alpha)$. Any desired $2\times 2$ unitary with a required splitting $\theta$ can then be implemented  by setting $\theta_\alpha$ such that $2\lvert \alpha(\theta_\alpha) + \beta \rvert < \theta < 2\lvert \alpha(\theta_\alpha) - \beta \rvert$ and correcting the resultant phase errors (Supplementary Info., Sec. III.B.).

Not all optical gates within the circuit necessarily need to incorporate redundancy. High accuracy unitary circuits have been demonstrated by incorporating only a few extra MZIs into the circuit, which can be trained using nonlinear optimization \cite{burgwal_using_2017} or gradient descent \cite{pai_matrix_2019}. Error correction serves an important purpose for these circuits, as one can optimize the hardware settings once on an ideal model and port the settings over to many devices.
For recirculating meshes the phase shifter settings are not constrained by the Haar measure, and so the benefit gained from error correction is not expected to diminish with increasing $N$. We therefore expect error correction to play an important role in scaling up the size of these circuits as well.

The motivation for photonic error correction assumes the hardware is re-programmed infrequently, for instance to implement a weight matrix in a neural network. Other applications, such as mode unscrambling, require real-time configuration robust to device error. We have recently discussed error-resilient self-configuration approaches in \cite{hamerly_stability_2021, hamerly_accurate_2021}.

\section{Conclusion}
In conclusion, we have presented a protocol to  correct for hardware errors in programmable photonic circuits. Unlike optimization-based approaches, our protocol utilizes a one-time calibration procedure to flexibly implement any desired functionality up to the limits of the hardware. We find that applying our approach to key application areas of programmable photonics, such as optical neural networks and programmable coupled-ring systems, enables resilience to fabrication errors well beyond modern-day process tolerances. Error correction also greatly reduces the overhead for programmable photonics that require optimization to deduce the hardware settings, as it eliminates the need to retrain for each individual set of hardware with unknown fabrication errors. Current process tolerances suggest that our approach enables improved functionality for  systems of up to hundreds of modes, providing a new avenue for scaling up programmable photonics.

\begin{backmatter}
  \bmsection{Funding} National Science Foundation (NSF) Graduate Research Fellowship Program (1745302); Air Force Office of Scientific Research (AFOSR) (FA9550-20-1-0113; FA9550-16-1-0391); IC Postdoctoral Research Fellowship Program.

  \bmsection{Acknowledgments} The authors are grateful to Hugo Larocque and Alexander Sludds for helpful comments on the manuscript.

  \bmsection{Disclosures} The authors are inventors on a provisional patent application on error correction methods for programmable photonics.

  \bmsection{Data availability} The data that support the plots in this paper are available from the corresponding author upon reasonable request.

  \bmsection{Supplemental document}
  See Supplement 1 for supporting content. 
\end{backmatter}

\bibliography{main}

\end{document}


\maketitle

\section{Hardware Calibration}

Error correction requires a one-time calibration of each phase shifter and passive splitter in the photonic circuit. While characterization of the overall linear transformation $U$ performed by a circuit is fairly straightforward \cite{rahimi-keshari_direct_2013}, the lack of direct access to individual optical elements makes measurement of their characteristics quite challenging. In order to enable our approach to hardware error correction, we have developed a protocol to calibrate all components on chip with simple interference measurements and homodyne detection on the circuit outputs. Importantly, our approach yields the circuit parameters directly from the measurements and does not rely on detectors embedded within the circuit. In this section, we illustrate our protocol for a rectangular (Clements) unitary circuit \cite{clements_optimal_2016}; however, our approach can be readily applied to any arbitrary network of MZIs.

A key consideration for our protocol is that the finite extinction ratio of the MZIs causes small amounts of spurious, unwanted light to scatter randomly into each device. A device is usually characterized by inputting an optical signal $E_1$ into one device port and measuring the transmission as $\theta, \phi$ are varied. However, characterizing a device in the middle of the network requires routing the optical signal through other interferometers, each of which are programmed to cross ($\theta=0$) or bar ($\theta=\pi$) configurations to route the light through wire paths within the circuit (Fig.~\ref{calibrate}a). If the routing MZIs are ideal, then cross or bar settings will direct the input probe signal $E_1$ into the desired output. 

Imperfect devices, however, are unable to realize ideal cross or bar configurations; as a result, a small amount of light $\sin(\alpha \pm \beta) E_1$ exits the unwanted output wherever light interacts with a device in the optical path. These spurious signals scatter randomly throughout the network; as a result, any device being characterized with a signal $E_1$ into one input will also have a small, unwanted signal $E_2 = \zeta E_1$ ($\zeta \ll 1)$ incident upon the other input. If unaccounted for, this extraneous light introduces an error $O(\zeta)$ in the calibration protocol which can be on the order of the beamsplitter errors being corrected.

One strategy for isolating the probe signal during characterization is to apply a low-frequency modulation to the devices the signal is traveling through \cite{mower_high-fidelity_2015}, which enables spurious light traveling through unwanted devices to be eliminated in the Fourier transform of the output. In this section, we present an alternate strategy applicable to devices where the relative phase of the two inputs are controllable with a phase shifter.

\subsection{Device Calibration}

\begin{figure*}[tb]
    \centering
    \includegraphics[width=\textwidth]{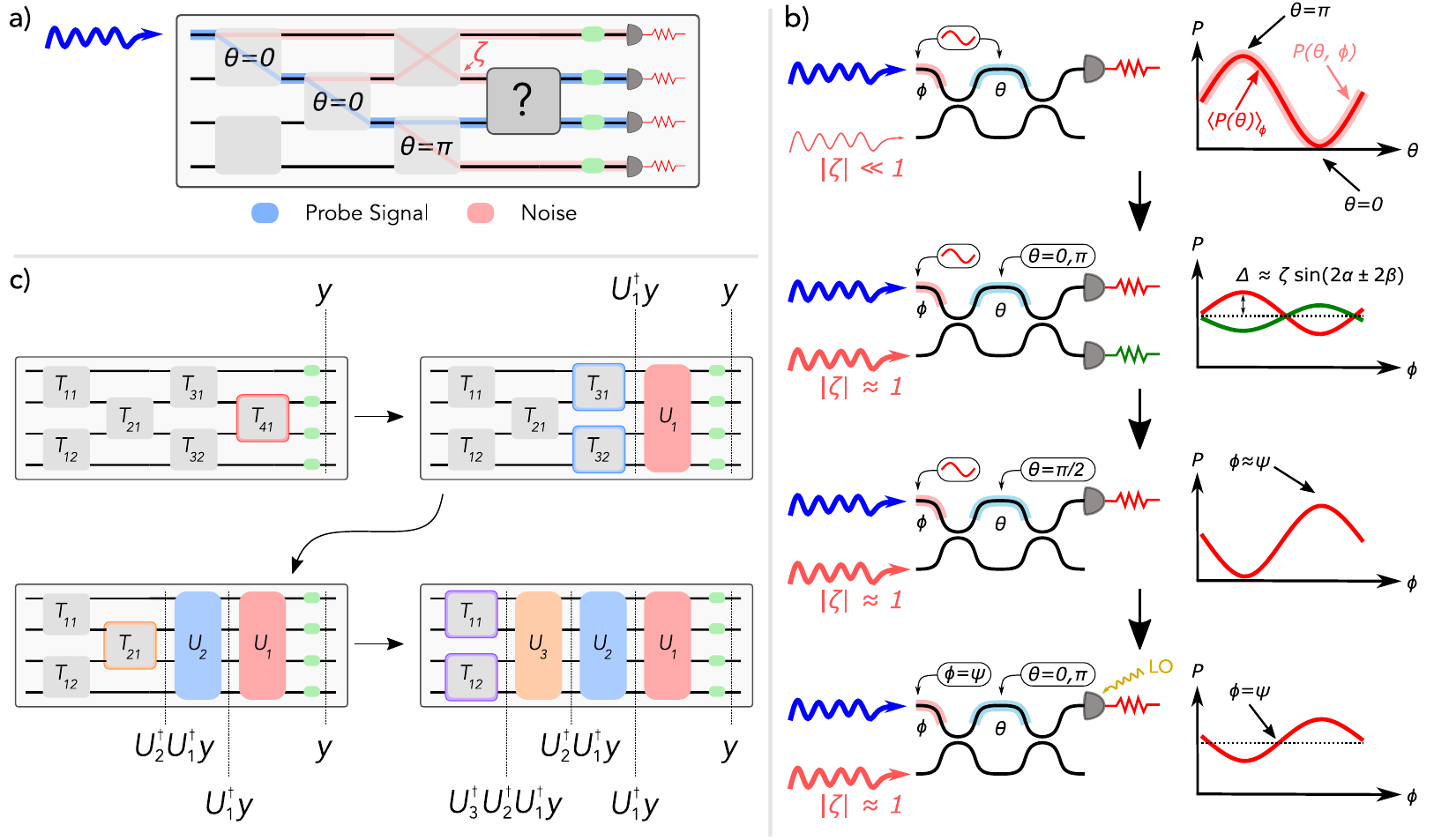}
    \caption{a) Devices are characterized by programming ``wire paths'' into the circuit, where each MZI along the route is set to the cross or bar state. However, device errors along the path scatter spurious light throughout the circuit, introducing errors into calibration. b) Device calibration is conducted in four steps. First, an optical signal is sent into one port of the MZI. Optimizing the output optical power averaged over $\phi$ produces an accurate calibration for $\theta$. The second step is to input optical signals into both ports and modulate the phase of one input while setting $\theta$ to the cross and bar states to measure $|\alpha + \beta|, |\alpha-\beta|$. This procedure has a sign ambiguity that is resolved in the third step, where the MZI is programmed to act as a 50-50 beamsplitter and output power vs. $\phi$ is optimized to set the two inputs into phase. Resetting to the cross/bar states and driving $\phi$ then allows us to deduce  the signs of $\alpha, \beta$. The final step is to set the two inputs into phase and use coherent detection to measure the difference in the phase of the output field between cross and bar states, which provides a calibration for $\phi$. c) Circuit calibration is completed from the output working backwards to the input. For devices in the middle of the network homodyne detection can be used to infer the output fields at any device of interest.}
    \label{calibrate}
\end{figure*}

The procedure for calibrating a single MZI is sketched in Figure \ref{calibrate}b. Calibration does not require internal detectors at each MZI, but does assume coherent detection at the outputs capable of reconstructing the field amplitude and phase. 

We start with devices connected directly to the detectors at the circuit output, which in a rectangular unitary circuit corresponds to the devices in the last column. Light is input into the circuit and each MZI within the path is optimized to maximize the signal input into the device of interest. The precise amount of light incident upon the device being characterized is not important, only that enough light reaches the MZI to produce measurements above the detector noise floor. We assume the input signal vector to the MZI is of the form $\mathbf{x} = E[1, \zeta e^{i\psi}]$, where $E$ is not known and $\zeta \ll 1$ is an unknown scaling factor indicating the amount of spurious light entering the other input.

$\bm{\theta}$ \textbf{calibration}: The first step is to calibrate the internal phase shifter $\theta$ by sweeping the voltage applied to the phase shifter and measuring the output transmission. We input a strong signal into the top input of the device; due to device errors, a small, unknown input with relative phase $\psi$ and amplitude $|\zeta| \ll 1$ will also be incident upon the bottom input. The power exiting the top output port is proportional to:
\begin{align*}
    P_\text{top} = \underbrace{\frac{1}{2} \left ( 1 + |\zeta|^2 + \left ( 1 - |\zeta|^2 \right ) \sin (2\alpha) \sin (2\beta) \right)}_\text{constant} &+ \underbrace{\frac{1}{2}(|\zeta|^2-1) \cos (2\alpha) \cos (2\beta) \cos \theta}_\theta - \underbrace{|\zeta| \cos (2\alpha) \sin (2\beta) \sin(\phi-\psi)}_{\zeta, \phi} + \\ & \!\!\!\!\!\!\!\! \underbrace{|\zeta| \cos (2\beta) \left ( \cos^2 \left ( \frac{\pi}{4}+\alpha \right ) \sin(\theta+\phi-\psi) + \sin^2 \left ( \frac{\pi}{4}+\alpha \right ) \sin(\theta-\phi+\psi) \right )}_{\zeta, \theta,\phi} \numberthis
\end{align*}

If $|\zeta| = 0$, we can calibrate $\theta$ by observing that $P_\text{top}$ is minimized at $\theta=0$. However, if $|\zeta| \neq 0$, there are contributions to $P_\text{top}$ dependent solely on $\phi$ and jointly on both $\theta, \phi$; as a result, optical power is minimized at:
\begin{equation}
    \theta_\text{max} = \arctan \left [ \frac{2 |\zeta| \cos (\phi - \psi)}{(|\zeta|^2-1)\cos (2\alpha) - 2|\zeta| \sin (2\alpha) \sin(\phi-\psi)} \right ] \approx -2 |\zeta| \cos (\phi - \psi)
\end{equation}

Simply optimizing $P_\text{top}$ with respect to $\theta$ would therefore produce a calibration error on the order of $O(\zeta)$. 

We can get around this by observing that if we average $P_\text{top}$ over all values of $\phi$, this measurement is always minimized at $\theta = 0$:
\begin{equation}
    \langle P_\text{top} \rangle_\phi = \frac{1}{2} \left ( 1 + |\zeta|^2 + \left ( 1 - |\zeta|^2 \right ) \sin (2\alpha) \sin (2\beta) \right) + \frac{1}{2}(|\zeta|^2-1) \cos (2\alpha) \cos (2\beta) \cos \theta
\end{equation}
$\theta$ is therefore calibrated by constructing the two-dimensional transmission characteristic $P_\text{top}(\theta, \phi)$ and optimizing the average transmission over all settings for $\phi$. The phase shifter setting for $\theta = \pi$ can similarly be obtained by maximizing this measurement, and arbitrary phase settings can be found by fitting this expression to the  measured transmission.

$\bm{\alpha,\beta}$ \textbf{calibration}: We can now use this information to program with high fidelity the bar ($\theta = \pi$) or cross ($\theta = 0$) settings into the MZI. These settings correspond to the unitaries:
\begin{equation} U_\text{bar} =
  - \left[ {\begin{array}{cc}
  e^{i \phi} \cos(\alpha - \beta) &  i \sin (\alpha - \beta) \\
    -e^{i \phi} i \sin (\alpha - \beta) &  - \cos(\alpha - \beta)
  \end{array} } \right]
\end{equation}
\begin{equation} U_\text{cross} =
    i \left[ {\begin{array}{cc}
     e^{i \phi} i \sin (\alpha + \beta) &  \cos(\alpha + \beta) \\
      e^{i \phi} \cos(\alpha + \beta) & i \sin (\alpha + \beta)
    \end{array} } \right]
\end{equation}
For ideal devices $\alpha=\beta=0$ these unitaries reduce to identity and swap operations, respectively. 

The beamsplitter calibration is now performed by sending roughly equal amounts of light into both inputs, i.e. applying an input field vector $\mathbf{x} = E[1, \zeta e^{i\psi}]$ where $\zeta \approx 1$ but once more the precise scaling factor is unknown. This can be achieved by either inputting coherent light into two inputs of the circuit, or by inputting light into one port and programming an MZI earlier along the wire path to operate as an approximate 50-50 beamsplitter. 

We first set $\theta = 0$ and measure the photocurrent $I_\text{top}, I_\text{bottom}$ at both outputs as a function of the external heater $\phi$:
\begin{equation}
    I_{\text{top}, \theta=0} = R_\text{top} |E|^2 (1 + (|\zeta|^2-1)\cos^2(\alpha+\beta) - |\zeta| \sin(2(\alpha+\beta)) \sin (\phi - \psi))
    \label{Itop}
\end{equation}
\begin{equation}
    I_{\text{bottom}, \theta=0} = R_\text{bottom} |E|^2 (1 + (|\zeta|^2-1)\sin^2(\alpha+\beta) + |\zeta| \sin(2(\alpha+\beta)) \sin (\phi - \psi))
    \label{Ibot}
\end{equation}
where $R_\text{top}, R_\text{bottom}$ are the unknown responsivities of the photodetectors.
This measurement produces a modulation of the photocurrent as the relative phase $\phi - \psi$ between inputs (controlled by $\phi$) is varied. The interference visibilities $\Delta = (I_\text{max}-I_\text{min})/(I_\text{max}+I_\text{min})$ for the top and bottom outputs are:
\begin{equation}
    \Delta_{\text{top}, \theta=0} = \frac{|\zeta \sin (2(\alpha+\beta))|}{1 + (|\zeta|^2 - 1) \cos^2(\alpha+\beta)}
\end{equation}
\begin{equation}
    \Delta_{\text{bottom}, \theta=0} = \frac{|\zeta \sin (2(\alpha+\beta))|}{1 + (|\zeta|^2 - 1) \sin^2(\alpha+\beta)}
\end{equation}
Solving this system of equations will yield values for $\zeta$ and $|\alpha+\beta|$.

Repeating this procedure for $\theta = \pi$ provides expressions that can be solved to find $|\alpha - \beta|$:
\begin{equation}
    \Delta_{\text{top}, \theta=\pi} = \frac{|\zeta \sin (2(\alpha-\beta))|}{1 + (|\zeta|^2 - 1) \sin^2(\alpha-\beta)}
\end{equation}
\begin{equation}
    \Delta_{\text{bottom}, \theta=\pi} = \frac{|\zeta \sin (2(\alpha-\beta))|}{1 + (|\zeta|^2 - 1) \cos^2(\alpha-\beta)}
\end{equation}
In the limit of $\zeta \rightarrow 1$, the interference visibilities are related directly to the beamsplitter errors, i.e. $\Delta_{\theta=0} = \sin (2(\alpha+\beta))$ and $\Delta_{\theta=\pi} = \sin (2(\alpha-\beta))$. 

This procedure characterizes how much the two input modes mix through interference when the MZI is set to the cross and bar states. In an ideal device, the bar and cross configurations implement identity and swap operations,  inhibiting interference between the input modes. Any observed interference is therefore the product of beamsplitter errors within the MZI. Inputting roughly equal amounts of light into both inputs ($\zeta \approx 1$) maximizes the interference visibility, which has the advantage of being insensitive to detector responsivity and out-coupling loss.

The final step is to deduce the sign of $\lvert \alpha+\beta \rvert$ and $\lvert \alpha-\beta\rvert$. To do this, we set $\theta = \pi/2$ and tune $\phi$ to maximize power exiting the top port, which occurs when $\phi \approx \psi$. Having identified the external phase shifter setting corresponding to $\psi$, we can now reset back to the cross state. If $I_\text{top}$ increases (decreases) when the phase shifter voltage is increased, then $\alpha + \beta$ is negative (positive). The procedure is repeated for the bar state to determine the sign of $\alpha - \beta$. These measurements provide sufficient information to compute $\alpha, \beta$ exactly.

$\bm{\phi}$ \textbf{calibration}: Precise calibration of $\phi$ requires measurement of the output field phase with coherent detectors. We input a strong optical signal into both ports, program the MZI to the cross state, and tune $\phi$ to precisely set $\phi = \psi$ (equations (\ref{Itop}), (\ref{Ibot}), after setting $\phi \approx \psi$ using the procedure above). The phase of the signal exiting the top output is:
\begin{equation}
    \arg E_{\text{top}, \theta=0} = \psi + \mathrm{atan2} \left [ |\zeta| \cos (\alpha+\beta), -\sin(\alpha+\beta) \right ]
\end{equation}

Now set the MZI to the bar state and measure the output phase once more. We obtain:
\begin{equation}
    \arg E_{\text{top}, \theta=\pi} = \psi + \mathrm{atan2} \left [ -|\zeta| \sin (\alpha-\beta), -\cos(\alpha-\beta) \right ]
\end{equation}

Solving this system of equations provides  $\psi$; using this information, we can now program any arbitrary phase $\phi$.

\subsection{System Calibration}
This procedure can be used to characterize $\theta, \phi, \alpha, \beta$ for all MZIs connected directly to photodetectors. We can therefore directly obtain the unitary $U_1$ corresponding to the last column of the circuit (Fig.~\ref{calibrate}c). With this information, we can now directly obtain the fields exiting an MZI in the preceding column $U_2$ by using homodyne detection to reconstruct the output field amplitude vector $\mathbf{y}$; the fields exiting an MZI in the penultimate column can be back-calculated to be $U_1^\dagger \mathbf{y}$.

The characterization therefore proceeds one column at a time, starting from the output side and working backwards towards the input. Homodyne detection allows direct measurement of fields exiting any MZI in the network; for an MZI in column $k$, the fields exiting that column will be $\prod_{k-1}^1 U^\dagger_i \mathbf{y}$. Each device is calibrated as before; however, instead of directly measuring photocurrents $I_1, I_2$, the output fields at each device are inferred with homodyne measurements of $\mathbf{y}$. 

While we have illustrated the calibration protocol with a rectangular mesh, our approach can be applied to any arbitrary network of MZIs. The generalized procedure is to first characterize all devices directly connected to the output detectors. Using this information, MZIs one device removed from the outputs can then be characterized. This enables calibration of MZIs two devices removed from the outputs, and so on, until all devices within the circuit are characterized.

\subsection{Triangular Network Calibration}

This approach can be applied to any general network of interferometers. However, the symmetry of triangular (Reck) circuits enables a greatly simplified calibration procedure described here. In particular, only direct detection of intensities is required to calibrate the Reck circuit, rather than homodyne detection.

\begin{figure*}[tb]
    \centering
    \includegraphics[width=\textwidth]{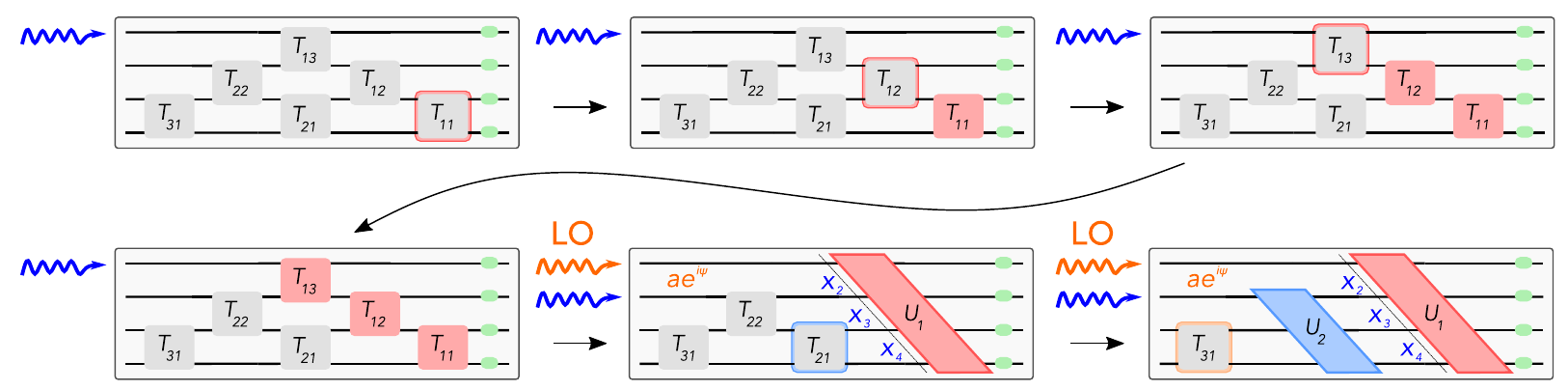}
    \caption{The simplified calibration procedure for a $4 \times 4$ Reck (triangular) circuit. Since each device is located along a diagonal, we can guarantee $\zeta = 0$ and extract $\alpha, \beta$ from direct extinction ratio measurements. Once the first diagonal $U_1$ is calibrated, we can program it to operate as an effective homodyne detector. Any other device in the circuit can be calibrated by interfering the output with $U_1$.}
    \label{reck_calibrate}
\end{figure*} 

The simplified procedure is depicted in Figure \ref{reck_calibrate}. Each diagonal of the circuit can be divided into sub-blocks $U_1, U_2,...,U_N$ which are characterized in  order. For each block, the MZIs are characterized starting at the end of the chain and working backwards.

The first device $T_{11}$ is characterized by inputting light into the first port of the circuit. Both outputs are directly connected to detectors, and the triangular structure of the network ensures that no light scatters into the bottom input, i.e. $\zeta=0$. This simplifies the procedure in several aspects:
\begin{itemize}
    \item The phase shifter $\theta$ can be calibrated by directly optimizing transmission vs. $\theta$, rather than having to first average transmission over $\phi$.
    \item Sweeping transmission vs. $\theta$ and computing the extinction ratio for the bar and cross ports gives the following expressions, which can be directly solved to find $\lvert \alpha\pm \beta \rvert$:
    \begin{equation}
        \text{ER}_\text{top} = \frac{I_{\text{top},\text{max}}}{I_{\text{top},\text{min}}} = \frac{\cos^2(\alpha-\beta)}{\sin^2(\alpha+\beta)}
    \end{equation}
    \begin{equation}
        \text{ER}_\text{bottom} = \frac{I_{\text{bottom},\text{max}}}{I_{\text{bottom},\text{min}}} = \frac{\cos^2(\alpha+\beta)}{\sin^2(\alpha-\beta)}
    \end{equation}
\end{itemize}
The signs of $\alpha, \beta$ can be determined interferometrically with the same approach as used in the generalized protocol. 

The second MZI characterized $T_{12}$ has the top port directly connected to a detector, while the output fields of the bottom port are determined by undoing the known operation $T_{11}^\dagger$. For the third device $T_{13}$, the fields exiting the bottom port are computed using $T_{11}^\dagger, T_{12}^\dagger$, and so on for the first diagonal.

Once the first diagonal is characterized, it can be programmed as a homodyne detector for the remainder of the circuit calibration. This is achieved by inputting a local oscillator field $ae^{i \psi}$ into the first port and programming $U_1$ to distribute equal power to all of the MZIs. Suppose we wish to measure the fields $x_2, x_3, ..., x_N$ exiting $U_2$.
Upon programming $U_1$, the fields exiting the circuit are $U_1(\mathbf{a} + \mathbf{x}) = U_1( [ae^{i \psi}, 0, 0, ..., 0]^T + [0, x_2, x_3, ..., x_N]^T)$. Since $U_1$ is programmed to distribute the LO signal equally to all outputs $y_i$, i.e. $U_1 \mathbf{a} = (ae^{i \psi}/\sqrt{N}) [1, 1, ..., 1]^T$, the field intensity $|y_i|^2$ at any port $i$ will be:
\begin{equation}
    |y_i|^2 = \frac{|a|^2}{N} + |U_1 \mathbf{x}|_i^2 + \frac{2a}{\sqrt{N}}~\mathrm{Re}[e^{-i \psi}U_1 \mathbf{x}]_i
\end{equation}
Taking measurements at $\psi=0, \pi/2$ will extract the in-phase and quadrature components of $U_1 \mathbf{x}$. This approach enables measurement of field amplitudes anywhere within the circuit; using it, we can characterize the remainder of the circuit $U_2, U_3, ..., U_N$ with intensity measurements only.

\section{Hardware Errors}

\subsection{Beamsplitter Errors}
The hardware error $\epsilon$ between a desired unitary matrix $U$ and the implemented matrix $U_\text{hardware}$ can be quantified by the Frobenius norm:
\begin{equation}
    \epsilon = \frac{1}{\sqrt{N}} \left (\sum_{ij} | U_{\text{hardware}, ij} - U_{ij}|^2 \right )^{1/2}
\end{equation}
This metric, which is bounded $\epsilon \in [0, 2]$, can be interpreted as an average relative error per entry of the matrix $U$; for example, in a neural network $\epsilon$ would correspond to the average relative error per weight. 

Unitary circuits decompose arbitrary matrices into a product of unitary matrices $T_{ij}(\theta,\phi,\alpha,\beta)$:
\begin{equation}
    U = D\prod_{ij} T_{ij}(\theta, \phi, \alpha, \beta)
\end{equation}
where $T_{ij}(\theta,\phi,\alpha,\beta)$ is:
\begin{equation}
\underbrace{
\begin{bmatrix}
    1 & 0 & \hdots & \hdots & \hdots & \hdots & 0 & 0\\
    0 & 1 & \hdots & \hdots & \hdots & \hdots &  0 & 0\\
    \vdots & & \ddots & & & \iddots & & \vdots \\
    \vdots & & & e^{i\theta} \cos\left ( \frac{\pi}{4} + \beta \right ) & i \sin\left ( \frac{\pi}{4} + \beta \right ) & &  & \vdots \\
    \vdots & & & i e^{i\theta} \sin\left ( \frac{\pi}{4} + \beta \right ) & \cos\left ( \frac{\pi}{4} + \beta \right ) & & & \vdots \\
    \vdots & & \iddots & &  & \ddots & & \vdots \\
    0 & 0 &&&&& 1 & 0 \\
    0 & 0 & \hdots & \hdots& \hdots& \hdots& 0 & 1
\end{bmatrix}}_{H_{2, ij}(\theta, \beta)}
\underbrace{
\begin{bmatrix}
    1 & 0 & \hdots & \hdots & \hdots & \hdots & 0 & 0\\
    0 & 1 & \hdots & \hdots & \hdots & \hdots &  0 & 0\\
    \vdots & & \ddots & & & \iddots & & \vdots \\
    \vdots & & & e^{i\phi} \cos \left ( \frac{\pi}{4} + \alpha \right ) & i \sin\left ( \frac{\pi}{4} + \alpha \right ) & &  & \vdots \\
    \vdots & & & i e^{i\phi} \sin\left ( \frac{\pi}{4} + \alpha \right ) & \cos\left ( \frac{\pi}{4} + \alpha \right ) & & & \vdots \\
    \vdots & & \iddots & &  & \ddots & & \vdots \\
    0 & 0 &&&&& 1 & 0 \\
    0 & 0 & \hdots & \hdots& \hdots& \hdots& 0 & 1
\end{bmatrix}}_{H_{1, ij}(\phi, \alpha)}
\end{equation}

The matrix error induced by a single beamsplitter error $\alpha$ can be computed as:
\begin{align}
    \epsilon = \frac{1}{\sqrt{N}} \left (\sum_{ij} |T_{ij}(\theta, \phi, \alpha=0, \beta=0) - T_{ij}(\theta, \phi, \alpha, \beta=0)|^2 \right )^{1/2}
\end{align}
The Frobenius norm is unitarily invariant, which originates from the cylic property of the trace; thus, only the unitary matrix corresponding to the beamsplitter error needs to be considered in the calculation of $\epsilon$:
\begin{align}
    \epsilon^2(\alpha) &= \frac{1}{N} \sum_{ij} |H_{1, ij}(\phi, \alpha) - H_{1, ij}(\phi, 0)|^2\\
    &=\frac{1}{N} \sum_{ij} \mathrm{Tr} \left [ (H_{1, ij}(\phi, \alpha) - H_{1, ij}(\phi, 0))^\dagger (H_{1, ij}(\phi, \alpha) - H_{1, ij}(\phi, 0)) \right ]\\
    &= \frac{1}{N} \mathrm{Tr} \left [ 2I - H_{1, ij}(\phi, \alpha)^\dagger H_{1, ij}(\phi, 0) - H_{1, ij}(\phi, 0)^\dagger H_{1, ij}(\phi, \alpha) \right ] \\
    &= \frac{1}{N} \left ( 2N - 2 \mathrm{Re} \left [ \mathrm{Tr} \left [ H_{1, ij}(\phi, \alpha)^\dagger H_{1, ij}(\phi, 0) \right ]\right ] \right ) \\
    &= \frac{1}{N} \left ( 2N - 2  \left ( 2\cos \alpha + N-2 \right ) \right ) \\
    &= \frac{4}{N} \left ( 1 - \cos \alpha \right ) \approx \frac{2 \alpha^2}{N}
\end{align}
Repeating this calculation for $\beta$ yields the same result. 

\begin{figure*}[tb]
    \centering
    \includegraphics[width=\textwidth]{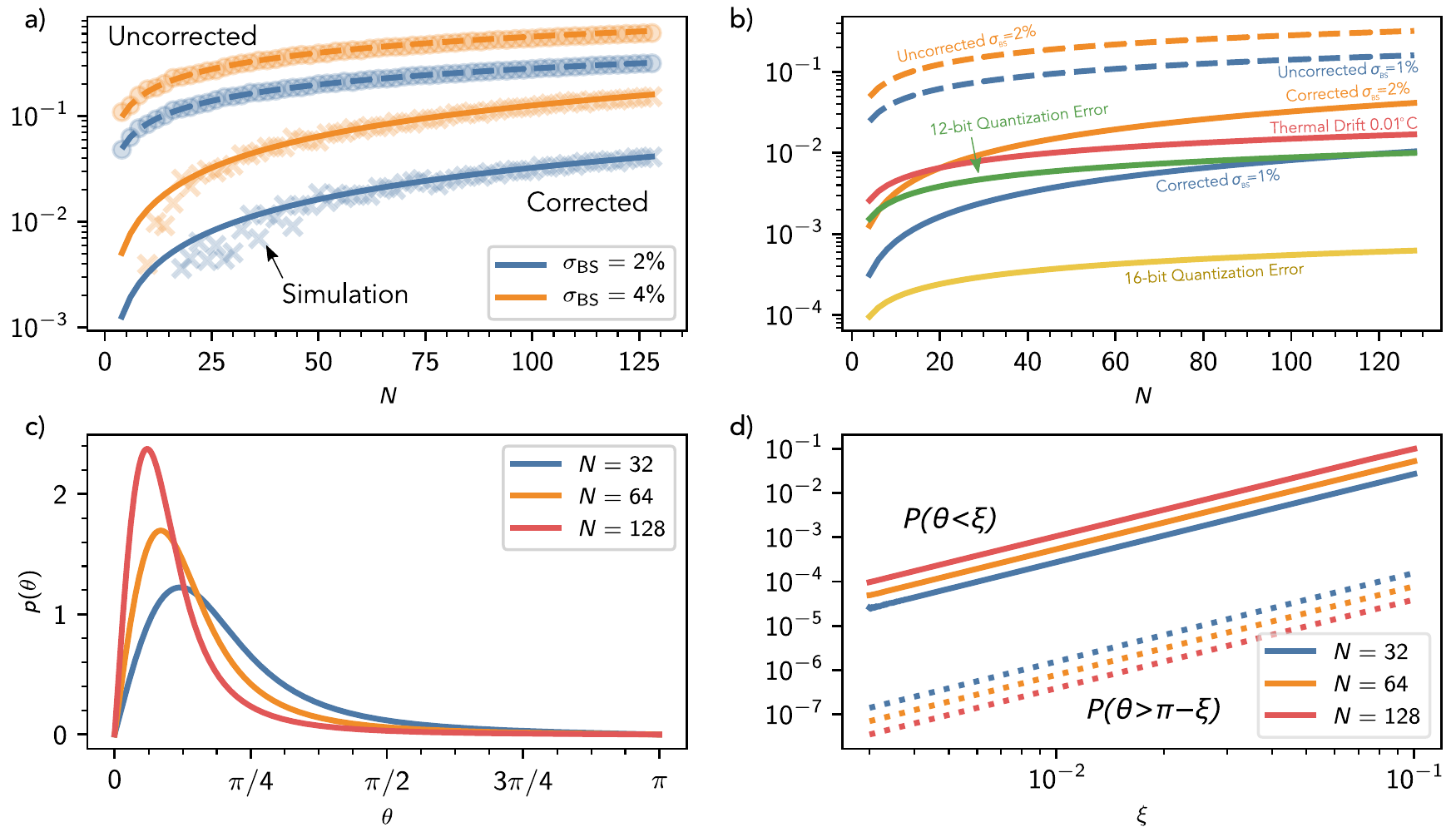}
    \caption{a) Equations (\ref{hardware_error}) and (\ref{corrected_hardware_error}) for the uncorrected and corrected beamsplitter errors as a function of circuit size $N$. The scatter plot shows the median error for 12 simulations, showing excellent agreement with the derived expressions. b) The relative error contributions from beamsplitter error, thermal drift, and quantization error as a function of circuit size $N$. If the component errors are left uncorrected, then even small beamsplitter variations produce errors significantly larger than those produced by dynamic effects. Hardware error correction suppresses these component errors to a point where dynamic effects begin to play an important role, particularly if the DAC resolution is low. c) The probability density function of the internal phase shifter setting $\theta$ for $N = \{32, 64, 128\}$. As $N$ increases, $\langle \theta \rangle$ is further biased towards $0$. d) The probability an MZI must be programmed to a splitting $\theta < \xi$, $\theta > \pi - \xi$ for $N = \{32, 64, 128\}$. $P(\theta>\pi - \xi)$ is orders of magnitude smaller than $P(\theta < \xi)$; thus, we can neglect it when computing the expected corrected hardware error.}
    \label{errors}
\end{figure*} 

In a unitary circuit with $N(N-1)/2$ interferometers, the average error is therefore:
\begin{align}
    \langle \epsilon \rangle &= \sqrt{\frac{N(N-1)}{2} \left (\langle \epsilon^2(\alpha) \rangle + \langle \epsilon^2(\beta) \rangle \right )} \\
    &= \sqrt{\left (N-1 \right ) \left ( \langle \alpha^2 \rangle + \langle \beta^2 \rangle \right )}\\ 
    &= \sqrt{2 \left (N-1 \right )}\sigma_\text{BS} \numberthis \label{hardware_error}
\end{align}

Figure \ref{errors}a shows the expression in Equation (\ref{hardware_error}) plotted against simulation results; they show excellent agreement with the derived expression. 

\subsection{Phase Errors}
Our analysis in the main text focuses primarily on beamsplitter errors. There can also be errors in the phase shifter settings; however, the primary source of these errors is a static error originating from microscopic changes in waveguide geometry between the interferometer arms \cite{yang_phase_2015}. This static error is calibrated out in the first step of the characterization protocol. 

This calibration cannot account for dynamic errors, however. Potential sources of dynamic phase errors include thermal drift, thermal crosstalk between phase shifters, and quantization error. In this section, we show that the contribution of these effects to the hardware error is significantly smaller than the static errors considered in the main text.

To start, we find that any error $\Delta$ induced in a single phase setting by these effects can be computed to be:
\begin{align}
    \epsilon^2(\Delta) &= \frac{1}{N} \left ( 2N - 2 \mathrm{Re} \left [ \mathrm{Tr} \left [ H_{2, ij}(\theta + \Delta , 0)^\dagger H_{2, ij}(\theta, 0) \right ]\right ] \right ) \\
    &= \frac{1}{N} \left ( 2N - 2 \left (\cos \Delta + N - 1  \right ) \right) \\
    &= \frac{1}{N} \left ( 2 - 2 \cos \Delta \right ) \\
    &\approx \frac{\Delta^2}{N}
\end{align}
We now consider the error induced by each of these effects.

\textbf{Thermal drift}: Typical thermo-electric cooling (TEC) systems can maintain chip temperature stabilities better than $<0.01^{\circ}$ C \cite{zhang_designing_2002}. The thermo-optic coefficient $\mathrm{d}n/\mathrm{d}T$ of silicon is $1.8 \times 10^{-4}~\mathrm{K}^{-1}$ \cite{komma_thermo-optic_2012}; for an $L = 200$ $\mu$m long phase shifter, a temperature gradient of $<0.01^{\circ}$ C therefore induces a phase error of $2 \pi (\mathrm{d}n/\mathrm{d}T) (\Delta T) L / \lambda \approx 1.5 \times 10^{-3}$ at $\lambda = 1550$ nm, which is an order of magnitude smaller than the expected beamsplitter error.

\textbf{Thermal crosstalk}: Thermal crosstalk is largely deterministic and dominated by the nearest-neighbor crosstalk, which can be accounted for in the phase shifter characterization. Additionally, crosstalk can be  suppressed by spacing interferometers sufficiently apart on the chip \cite{harris_efficient_2014}; a spacing of 135 $\mu$m, for instance, has been measured to generate a crosstalk with the neighboring MZI of less than 0.02 rad/rad \cite{jacques_optimization_2019}. Since thermal crosstalk decays with increasing separation, we expect with careful design this effect should not dominate hardware error.

\textbf{Quantization error}: Quantization error originates from the digital-to-analog converters (DACs) used to program voltages into the phase shifters. Consider an $N$-bit DAC whose $2^N$ codewords range from zero voltage to the voltage $V_{2\pi}$ required for a $2 \pi$ phase shift. Programming the $M$-th ($0 \leq M \leq 2^N-1$) codeword will produce a voltage sampled uniformly over the distribution:
\begin{equation} 
    V_M = \frac{V_{2\pi}}{2^N} \left ( M + \frac{1}{2} \right ) \pm \underbrace{\frac{V_{2\pi}}{2^{N+1}}}_{N~\text{bits}}
\end{equation}
In a thermo-optic phase shifter, relative phase is a function of the voltage squared; the phase setting for the $M$-th codeword is therefore:
\begin{align} 
    \phi_M &= \frac{2\pi}{2^{2N}} \left ( M + \frac{1}{2} \pm \frac{1}{2} \right )^2\\
    &\approx \frac{2\pi}{2^{2N}} \left ( M + \frac{1}{2} \right )^2 \pm \frac{2\pi}{2^{2N}}\left ( M + \frac{1}{2} \right )
\end{align}

The uncertainty in $\phi$ is maximum at $M = 2^N-1$, where the phase setting is:
\begin{equation}
    \phi \approx 2\pi \pm \underbrace{\frac{2\pi}{2^N}}_{N-1~\text{bits}}
\end{equation}
which is one fewer bit of accuracy than for the voltage setting. 

The square-law dependence of phase on voltage therefore results in an $N$-bit DAC setting the phase to roughly $N-1$ bits of accuracy. A 12-bit DAC will suppress worst-case quantization error per phase shifter to $\approx 9 \times 10^{-4}$, and 16 bits are sufficient to suppress error to below $6 \times 10^{-5}$. 

In Figure \ref{errors}b we plot the relative error contributions of these effects compared to static beamsplitter error. These estimates suggest that \textit{uncorrected} component imprecision dominates the hardware error in programmable photonic circuits. However, once component errors are corrected, dynamic effects play a more significant role in the total hardware error.

\subsection{Calculating the corrected error}

As discussed in the main text, if $\theta^\prime, \phi^\prime$ are realizable for all devices in a circuit, then $\epsilon_\text{corrected} = 0$. For large circuit sizes $N$, however, some devices will require a splitting $\theta$ outside the range of realizable values $2|\alpha+\beta|<\theta<\pi - 2|\alpha-\beta|$. 

Consider a device for which we can correct $\phi, \psi_1, \psi_2$, but are unable to correct $\theta$. Any unitary $U$ can be decomposed into a product of matrices $U = D \prod T_{ij}$, where $D$ is diagonal and $T_{ij}$ is a $N\times N$ block matrix with non-trivial entries:
\begin{equation}
   \begin{bmatrix} e^{i \psi_1} & 0 \\ 0 & e^{i \psi_2} \end{bmatrix} \begin{bmatrix} \sin (\theta/2) & \cos (\theta/2) \\ \cos (\theta/2) & -\sin (\theta/2) \end{bmatrix} 
   \begin{bmatrix} e^{i \phi} & 0 \\ 0 & 1 \end{bmatrix}
\end{equation}

An error $\theta \rightarrow \theta + \Delta$ produces a contribution to $\epsilon_\text{corrected}$ of:
\begin{align}
    \epsilon^2(\Delta) &= \frac{1}{N} \left (2N - 2(2 \cos(\Delta/2) + N-2) \right )\\
    &= \frac{8}{N} \sin^2(\Delta/4) \approx \frac{\Delta^2}{2N}
 \end{align}

On average, given $\theta$ cannot be realized, $\langle \Delta^2 \rangle = 2(\langle \alpha^2 \rangle + \langle \beta^2 \rangle) = 4\sigma_\text{BS}^2$ and the error per device will be $\langle \epsilon^2(\Delta) \rangle = 2\sigma_\text{BS}^2/N$. The total error for the circuit is therefore:
\begin{equation}
    \langle \epsilon_\text{corrected} \rangle = \sqrt{(N-1) \sigma_\text{BS}^2 P(\theta<2|\alpha+\beta|)}
\end{equation}
where $P(\theta<2|\alpha+\beta|)$ is the probability that a device in the circuit needs to be programmed to a splitting that cannot be realized.

The distribution of internal phase shifter settings $\theta$ for a unitary circuit can be determined from the Haar measure. For a given MZI, ref.~\cite{russell_direct_2017} shows that:
\begin{equation}
    p_{n,i}(\theta) = (n-i) \sin (\theta/2) \cos^{2(n-i)-1} (\theta/2)
\end{equation}
where $n \in [2, N], i \in [1, N - n + 1]$ are indices denoting the position of the MZI in the network (see \cite{russell_direct_2017} for the mapping). The distribution of $\theta$ over the entire circuit can therefore be written as (Fig.~\ref{errors}c):
\begin{align}
    p(\theta) &= \sum_{k=1}^{N-1} \frac{2(N-k)}{N(N-1)} k \sin (\theta/2) \cos^{2k-1} (\theta/2)
 \end{align}
Integrating this expression yields the fraction of beamsplitters with a required splitting below $\xi$:
\begin{align}
   P(\theta < \xi) &= \sum_{k=1}^{N-1} \frac{2(N-k)}{N(N-1)}  \int_{0}^\xi k \sin (\theta/2) \cos^{2k-1} (\theta/2)~\mathrm{d}\theta\\
   &=\sum_{k=1}^{N-1} \frac{2(N-k)}{N(N-1)} \left ( 1 - \cos^{2k}(\xi/2)\right ) \label{probability}\\
   &= \frac{N+1}{N-1} - \frac{4\left ( N +  \cot ^2\left(\xi/2 \right) \left(\cos ^{2 N}\left(\xi/2 \right)-1\right) \right )}{N(N-1)(1-\cos \xi)} 
\end{align}
For small device errors, equation (\ref{probability}) can be Taylor expanded to:
\begin{equation}
    \sum_{k=1}^{N-1} \frac{2(N-k)}{N(N-1)} \left (\frac{k \xi^2}{4} \right) = \frac{N+1}{12} \xi^2 = \frac{2(N+1)}{3} \sigma_{BS}^2
\end{equation}

On the other hand, the probability that $\theta > \pi - 2|\alpha - \beta|$ is:
\begin{align}
    P(\theta > \pi - 2|\alpha - \beta|) &= \sum_{k=1}^{N-1} \frac{2(N-k)}{N(N-1)}  \int_{\pi - 2|\alpha - \beta|}^\pi k \sin (\theta/2) \cos^{2k-1} (\theta/2)~\mathrm{d}\theta\\
    &=\sum_{k=1}^{N-1} \frac{2(N-k)}{N(N-1)}  \cos^{2k}\left (\frac{\pi}{2} - |\alpha - \beta| \right ) \\
    &\approx \sum_{k=1}^{N-1} \frac{2(N-k)}{N(N-1)} 2^k \sigma_{BS}^{2k} \approx \frac{4 \sigma_{\text{BS}}^2}{N}
\end{align}
For moderately large $N$, this quantity is order of magnitudes smaller  than $P(\theta < 2|\alpha+\beta|)$; we can therefore disregard it when estimating the average corrected error (Fig.~\ref{errors}d).

The average corrected error is therefore:
\begin{align}
    \langle \epsilon_\text{corrected} \rangle &= \sqrt{(N-1) \sigma_\text{BS}^2 P(\theta<2|\alpha+\beta|)}\\
    &= \sqrt{(N-1) \sigma_\text{BS}^2 \left (\frac{2(N+1)}{3} \sigma_{BS}^2 \right )} \\
    &= \sigma_\text{BS}^2 \sqrt{\frac{2(N^2-1)}{3}} \label{corrected_hardware_error}
\end{align}
This expression is plotted in Figure \ref{errors}a and also shows excellent agreement with simulation results.

\section{Hardware Error Correction}
\subsection{Correcting the internal phase shifter}
In this section, we derive equation (9) in the main text providing the correction to the internal phase shifter $\theta$ for an imperfect device. Programming an MZI with phase settings $(\theta, \phi)$ produces the unitary:
\begin{align}
    T_{ij}(\theta, \phi)
    &= i e^{i \theta/2} \begin{bmatrix} e^{i \phi} \sin (\theta/2) & \cos (\theta/2) \\ e^{i \phi} \cos (\theta/2) & - \sin (\theta/2) \end{bmatrix}
\end{align}

However, an MZI with splitting errors $\alpha, \beta$ implements the unitary $T^\prime_{ij}(\theta, \phi, \alpha, \beta)$:
\begin{align}
    T^\prime_{ij}(\theta, \phi, \alpha, \beta) 
     &= i e^{i \theta / 2} \begin{bmatrix}
        e^{i \phi} (\cos(\alpha - \beta) \sin \frac{\theta}{2} + i  \sin (\alpha + \beta) \cos \frac{\theta}{2}) & \cos(\alpha + \beta) \cos \frac{\theta}{2} + i \sin (\alpha - \beta) \sin \frac{\theta}{2} \\
        e^{i \phi} (\cos(\alpha + \beta) \cos \frac{\theta}{2} - i \sin (\alpha - \beta) \sin \frac{\theta}{2} ) & - \cos(\alpha - \beta) \sin \frac{\theta}{2} + i \sin (\alpha + \beta) \cos \frac{\theta}{2}
     \end{bmatrix}
\end{align}

The correction $ \theta \rightarrow \theta^\prime$ can be derived by requiring that the magnitude of the upper left entry of $T^\prime_{ij} (\theta^\prime, \phi^\prime, \alpha, \beta)$ equal that of $T_{ij} (\theta, \phi)$. For a $2 \times 2$ unitary matrix $U$, the unitarity condition $U U^\dagger = I$ implies that setting the magnitudes of one term in both matrices to be equal is sufficient to set the magnitudes of all terms in the matrices to be equal. This condition produces an expression relating $\theta^\prime$ to $\theta$:
\begin{equation}
    \cos^2 (\alpha-\beta) \sin^2 (\theta^\prime/2) + \sin^2(\alpha+\beta) \cos^2 (\theta^\prime/2) = \sin^2 (\theta/2)
\end{equation}
Solving for $\theta^\prime$, we find that:
\begin{align}
    \sin^2 (\theta^\prime/2) &= \frac{\sin^2 (\theta/2) - \sin^2 (\alpha + \beta)}{\cos^2(\alpha - \beta) - \sin^2(\alpha+\beta)} \\
    \theta^\prime &= 2 \arcsin \sqrt{\frac{\sin^2 (\theta/2) - \sin^2 (\alpha + \beta)}{\cos^2(\alpha - \beta) - \sin^2(\alpha+\beta)}}
\end{align}

Since $\alpha, \beta$ are small, the denominator of the expression for $\theta^\prime$ will always be positive. This expression therefore has a solution only when the numerator is positive, i.e. $\sin^2(\theta/2) > \sin^2(\alpha+\beta)$, and when the argument in the $\arcsin$ function is less than 1, i.e. $\sin^2 \theta/2 - \sin^2 (\alpha + \beta) < \cos^2(\alpha - \beta) - \sin^2(\alpha+\beta)$. These conditions yield the range over which $\theta$ is physically realizable:
\begin{equation}
2|\alpha + \beta| < \theta < \pi - 2 \lvert \alpha-\beta \rvert
\end{equation}

\subsection{Perfect optical gates with redundant devices}

Device imperfections limit the range of realizable $\theta$ values. For unitary circuits this results in a net increase of $\epsilon$ with $N$, even with error correction, as more MZIs cannot be programmed to the required splitting. For recirculating waveguide meshes these errors will degrade the fidelity of the bar and cross configurations used to route signals, which induces unwanted crosstalk between systems.

\begin{figure*}[tb]
    \centering
    \includegraphics[width=\textwidth]{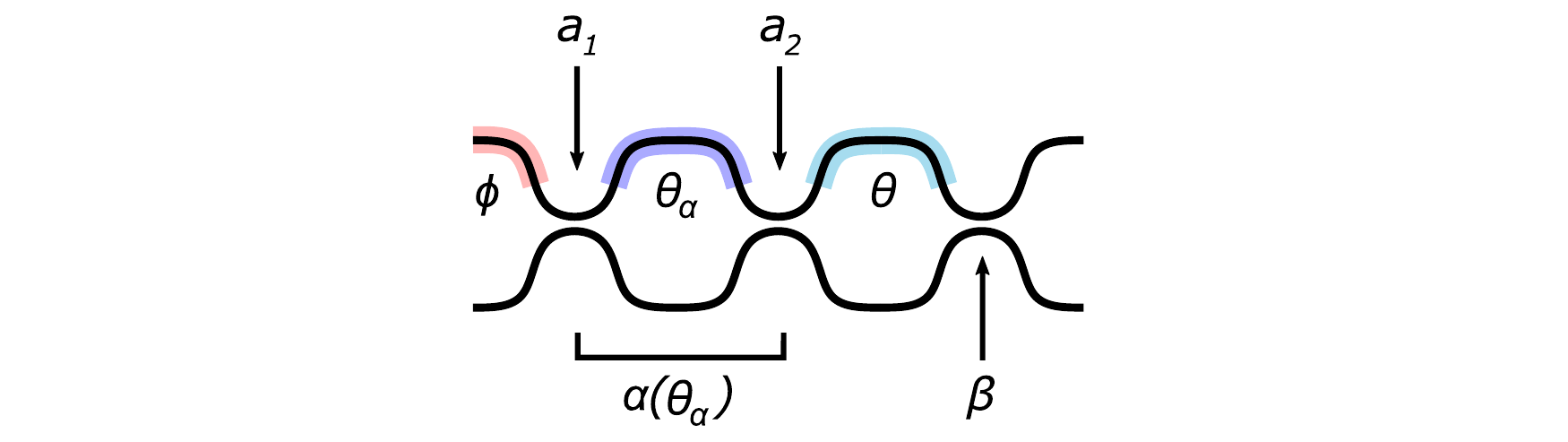}
    \caption{Redundant MZI for realizing perfect unitary optical gates. The splitters are assumed to have errors $a_1, a_2, \beta$.}
    \label{1p5mzi}
\end{figure*} 

An ideal optical gate can be realized with the redundant MZI shown in Figure \ref{1p5mzi} \cite{suzuki_ultra-high-extinction-ratio_2015, wang_tolerant_2020}. One beamsplitter is a passive splitter with error $\beta$, while the other is an MZI that implements a tunable error $\alpha(\theta_\alpha)$. The tunable splitter is assumed to consist of passive splitters with error $a_1, a_2$.

By setting $\alpha(\theta_\alpha) = -\beta$, we can implement any arbitrary splitting $0 \leq \theta \leq \pi - 4\beta$. Alternatively, we can set $\alpha(\theta_\alpha) = \beta$ to implement any desired splitting $4 \beta \leq \theta \leq \pi$. We can then correct for any phase errors using the usual procedure; thus, this interferometer can implement any arbitrary $2 \times 2$ unitary and a unitary circuit composed of these devices can always achieve $\epsilon_\text{corrected} = 0$.

A desired unitary $T_{ij}(\theta, \phi)$ could be obtained by many possible settings $(\theta^\prime, \phi^\prime, \theta_\alpha, \psi_1, \psi_2)$, since $\alpha(\theta_\alpha)$ is tunable over a wide range. One possible recipe for programming the device settings is:
\begin{itemize}
    \item If $\theta \lessgtr \pi/2$, set $\alpha(\theta_\alpha) = \mp \beta$. Following equation (1) in the main text, this requires programming $\theta$ to:
    \begin{equation}
         \theta_\alpha = 2 \arcsin \sqrt{\frac{\sin^2 (\pi/4 \pm \beta) - \sin^2 (a_1+a_2)}{\cos^2(a_1-a_2) - \sin^2(a_1+a_2)}}
    \end{equation}

    As long as $|\beta| < \pi/4 - \mathrm{max}[|a_1+a_2|, |a_1-a_2|]$, we can implement \textit{any} $2 \times 2$ unitary. If $a_1 = a_2 = \beta$, this constraint requires that the imbalance of each passive splitter is no larger than  75-25. This range is slightly smaller than the ``double MZIs'' proposed in \cite{miller_perfect_2015}, but will likely suffice for any foundry process and has the benefit of requiring one fewer phase shifter.
    \item If $\theta < \pi/2$, the optical transformation implemented for $\alpha(\theta_\alpha) = -\beta$ is:
    \begin{align}
        &ie^{i(\xi_3 + \theta_\alpha/2)} 
        \begin{bmatrix} \cos(\frac{\pi}{4} + \beta) & i \sin(\frac{\pi}{4} + \beta) \\ i \sin(\frac{\pi}{4} + \beta) & \cos(\frac{\pi}{4} + \beta) \end{bmatrix}
         \begin{bmatrix} e^{i (\theta^\prime+\xi_2-\xi_3)} & 0 \\ 0 & 1 \end{bmatrix}
         \begin{bmatrix} \sin(\frac{\pi}{4} + \beta) & \cos(\frac{\pi}{4} + \beta) \\ \cos(\frac{\pi}{4} + \beta) & -\sin(\frac{\pi}{4} + \beta) \end{bmatrix}
         \begin{bmatrix} e^{i (\phi^\prime + \xi_1 - \xi_2)} & 0 \\ 0 & 1 \end{bmatrix}\\
         = & -e^{i(\theta_\alpha + \theta^\prime +\xi_2  +\xi_3 - \pi/2)/2}
        \begin{bmatrix}
        e^{i\widetilde{\phi}} \cos (2\beta) \cos (\widetilde{\theta}/2) & -  \sin (2\beta) \cos (\widetilde{\theta}/2) + i \sin (\widetilde{\theta}/2)  \\ 
         i e^{i\widetilde{\phi}} \left ( \sin (2\beta) \cos (\widetilde{\theta}/2) + i \sin (\widetilde{\theta}/2) \right )
         &  i \cos (2\beta) \cos (\widetilde{\theta}/2)
        \end{bmatrix}
    \end{align}
    where:
    \begin{equation}
        (\widetilde{\theta},\ \widetilde{\phi})  = (\theta^\prime + \xi_2 - \xi_3 - \pi/2,\  \phi^\prime + \xi_1 - \xi_2)
    \end{equation}
    \begin{equation}
        (\xi_1,\ \xi_2,\ \xi_3) = \left ( \arctan \left [ \frac{\sin(a_1+a_2)}{\cos(a_1-a_2)} \cot \frac{\theta_\alpha}{2} \right ],\ \arctan \left [ \frac{\sin(a_1-a_2)}{\cos(a_1+a_2)} \tan \frac{\theta_\alpha}{2} \right ],\ -\arctan \left [ \frac{\sin(a_1+a_2)}{\cos(a_1-a_2)} \cot \frac{\theta_\alpha}{2} \right ] \right )
    \end{equation}

    We derive the settings to program $T(\theta, \phi)$ by following a procedure similar to that described in the main text. We find that:
    \begin{equation}
        \theta^\prime = 2 \arccos \left ( \sec (2\beta) \sin (\theta/2) \right ) + \frac{\pi}{2} + \xi_3 - \xi_2
        \label{1p5t1}
    \end{equation}
    \begin{equation}
        \phi^\prime = \phi - \xi_1 + \xi_2 + \arg \left [ -\sin(2\beta)\cos(\widetilde{\theta}/2) + i \sin (\widetilde{\theta}/2) \right ]
    \end{equation}
    \begin{equation}
        \psi_1 = \arg \left [ \sin(2\beta)\cos(\widetilde{\theta}/2) + i \sin(\widetilde{\theta}/2) \right ] + \frac{1}{2}\left(\theta - \theta_\alpha - \theta^\prime - \xi_2 - \xi_3 \right ) - \frac{5\pi}{4}
    \end{equation}
    \begin{equation}
        \psi_2 = \frac{\pi}{4} +\frac{1}{2}\left ( \theta - \theta_\alpha - \theta^\prime - \xi_2 - \xi_3 \right )
    \end{equation}
    \item If $\theta > \pi/2$, i.e $\alpha(\theta_\alpha) = \beta$, then the unitary transformation is instead:
    \begin{equation}
        -e^{i(\theta_\alpha + \theta^\prime +\xi_2  +\xi_3 - \pi/2)/2}
        \begin{bmatrix}
        e^{i\widetilde{\phi}} \left ( \cos (\widetilde{\theta}/2) - i\sin (2\beta) \sin (\widetilde{\theta}/2) \right ) & 
        i \cos (2\beta) \sin  (\widetilde{\theta}/2) \\ 
         -e^{i\widetilde{\phi}}\cos (2\beta) \sin  (\widetilde{\theta}/2)
         &  i \cos (\widetilde{\theta}/2) - \sin (2\beta) \sin (\widetilde{\theta}/2)
        \end{bmatrix}
    \end{equation}
    where $\widetilde{\theta}, \widetilde{\phi}, \xi_1, \xi_2, \xi_3$ are defined as earlier. In this case the required device settings are:
    \begin{equation}
        \theta^\prime = 2 \arcsin \left ( \sec (2\beta) \cos (\theta/2) \right ) + \frac{\pi}{2} + \xi_3 - \xi_2
        \label{1p5t2}
    \end{equation}
    \begin{equation}
        \phi^\prime = \phi - \xi_1 + \xi_2 + \frac{\pi}{2} + \arctan \left [ \tan (\widetilde{\theta}/2) \sin (2\beta) \right ]
    \end{equation}
    \begin{equation}
        \psi_1 = \frac{1}{2}\left (\theta - \theta_\alpha - \theta^\prime - \xi_2 - \xi_3 \right) - \frac{3\pi}{4}
    \end{equation}
    \begin{equation}
        \psi_2 = -\arg \left [ -\sin(2\beta)\sin(\widetilde{\theta}/2) + i\cos(\widetilde{\theta}/2) \right ] + \frac{1}{2}\left (\theta - \theta_\alpha - \theta^\prime - \xi_2 - \xi_3 \right) + \frac{3\pi}{4}
    \end{equation}
\end{itemize}

\begin{figure*}[tb]
    \centering
    \includegraphics[width=\textwidth]{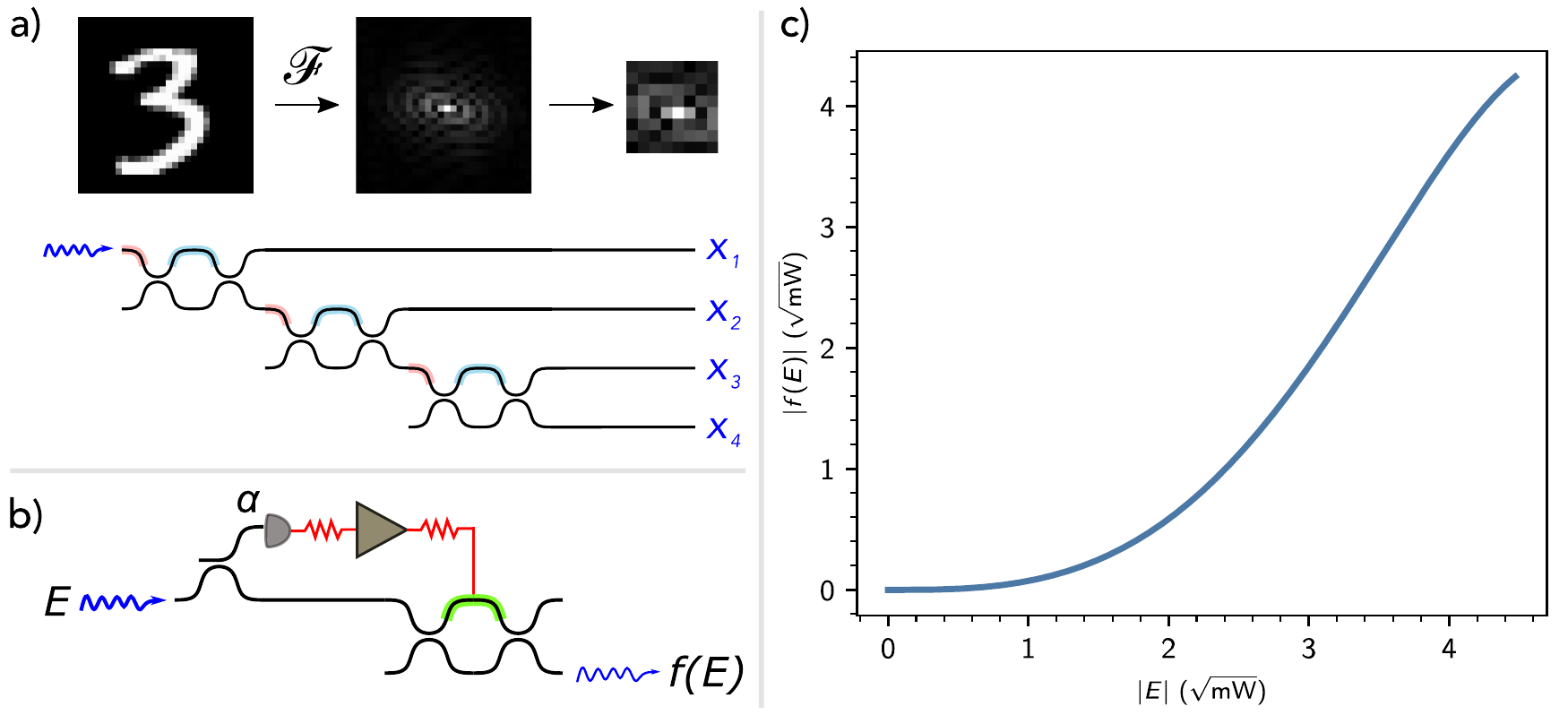}
    \caption{a) The MNIST data set was pre-processed with a Fourier transform and truncated to a $\sqrt{N} \times \sqrt{N}$ center window for a $N$-mode unitary circuit \cite{pai_parallel_2020}. The inputs were normalized to unit length, which can be realized optically with a diagonal line of MZIs. b) The activation function architecture as described in \cite{williamson_reprogrammable_2020}. A small fraction $\alpha$ of the input signal is tapped off to a photodiode driving a Mach-Zehnder modulator. c) The activation function $f(E)$ for the parameters used in the simulation. Since the hidden layers operate on electric field amplitudes, we plot the square root of the optical power in units $\sqrt{\text{mW}}$. Technically, $f(E)$ is non-monotonic for high optical powers, as the Mach-Zehnder interferometer will produce a $\cos(|E|^2)$ modulation. However, the input optical powers in our simulations are chosen to ensure the activation function operates only in the modReLU-like region.}
    \label{nn}
\end{figure*} 

\section{Methods}
The results presented in this paper were produced using a custom simulation package written in Python. The package simulates photonic circuits with the transfer matrix method and relies primarily on efficient array calculations with \verb|NumPy| \cite{harris_array_2020} and optimization routines included in \verb|SciPy| \cite{scipy_10_contributors_scipy_2020}. In this section, we provide further details on the application results presented in the main text.

\textbf{Optical neural networks}: The optical neural networks discussed in the main text were trained using the \verb|Neurophox| package. The neural network architecture, shown in Figure 4 of the main text, is based on the architecture described in \cite{pai_parallel_2020} with some modifications. Images of handwritten digits from the MNIST task are pre-processed with a Fourier transform and truncated to a $\sqrt{N} \times \sqrt{N}$ center window for a dimension $N$ unitary circuit. We assume a fixed amount of optical power is available to the circuit; each input vector corresponding to an image is normalized to unit length, so that all images are encoded into the neural network with the same amount of optical power. This normalization can be realized optically with a diagonal line of MZIs, as depicted in Figure \ref{nn}a.

The activation function is realized electro-optically with a tap photodiode coupled to a Mach-Zehnder modulator \cite{williamson_reprogrammable_2020} (Fig.~\ref{nn}b). The activation function taps off 10\% of the input power to the photodiode, while the remainder is directed to the modulator. The photocurrent drives the modulator through a transimpedance amplifier (TIA), resulting in a nonlinear modulation of the electric field. 

The nonlinearity implements the activation function \cite{williamson_reprogrammable_2020}:
\begin{equation}
    f(E) = (\sqrt{1-\alpha}) e^{-i(g|E|^2/2 + \phi/2 - \pi/2)} \cos ( g|E|^2/2 + \phi/2) E
\end{equation}
where $\alpha=0.1$ is the fractional power tapped off to the photodiode and $g = \pi/20$ is the modulator phase induced when 1 mW is incident upon the nonlinearity (prior to the tap). For typical electro-optic modulator drive voltages of $<8$ V \cite{streshinsky_silicon_2014, watts_low-voltage_2010} and a photodiode responsivity of 1 A/W \cite{zhang_high-responsivity_2014}, the required TIA gain for these parameters is roughly 36 dB$\Omega$. The modulator is biased so that no transmission occurs when $E=0$; as shown in Figure \ref{nn}c, for optical powers $<20$ mW $f(E)$ approximates a modReLU function \cite{arjovsky_unitary_2016}.

As the network size $N$ increases, the average power within a waveguide drops as $1/N$;  for this reason, we assumed the total optical power input into the circuit increased commensurately to ensure the activation function could still be triggered. The $N=\{36, 64\}$ networks were trained with 20 mW of optical power, the $N=144$ network was trained with 40 mW, and the $N=256$ network was trained with 60 mW of optical power. All of the neural networks were trained to minimize the mean squared error between the $L_2$ normalized output power and the one hot encoding of the correct image.

\textbf{Tunable dispersion compensators}: The tunable dispersion compensator was modeled with 15 serially coupled optical ring resonators implemented within a hexagonal waveguide mesh (Fig.~5; main text).

\begin{figure*}[t]
    \centering
    \includegraphics[width=\textwidth]{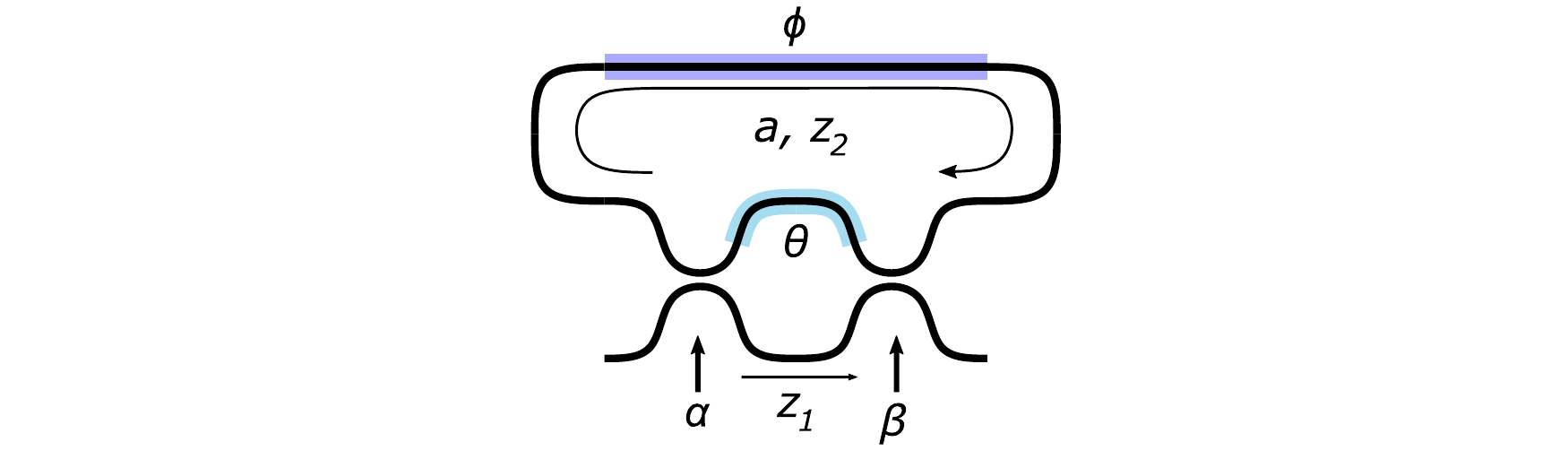}
    \caption{Model for tunable coupling ring. The ring coupling is set by an MZI with errors $\alpha, \beta$ and internal phase $\theta$, and the resonance is set with a phase setting $\phi$. The coupler is assumed to be lossless, and the feedback loop is assumed to have a round-trip transmission $a$.}
    \label{tunable_ring}
\end{figure*} 

The transfer function $T_i(\omega)$ of a single tunable coupling ring can be derived with Mason's gain formula \cite{mason_feedback_1953, mason_feedback_1956}:
\begin{equation}
    T_i(\omega) = \frac{a \cos^2(\alpha-\beta) e^{i(k(2z_1 + z_2) + \theta + \phi)} - \tau_1 \tau_2 e^{i k z_1} + \kappa_1 \kappa_2 e^{i(kz_1 + \theta)}}{a\tau_1 \tau_2  e^{i(k(z_1 + z_2) + \theta + \phi)} - a\kappa_1 \kappa_2 e^{i(k(z_1 + z_2) + \phi)} - 1}
\end{equation}
where $k = n(\omega) \omega/c$, $\tau_1 = \cos(\pi/4+\alpha)$, $\tau_2 = \cos(\pi/4+\beta)$, $\kappa_1 = \sin(\pi/4+\alpha)$, $\kappa_2 = \cos(\pi/4+\beta)$, $z_1$ is the interferometer arm length, $z_2$ is the length of the feedback loop, and $a$ is the round-trip transmission of the feedback loop (Fig.~\ref{tunable_ring}).

The transfer function $T_i(\omega)$ for each ring was individually computed and multiplied to yield the overall system response $T(\omega) = \prod_i T_i(\omega)$. From this result we found the group delay of the system $\tau(\omega) = -\mathrm{d}/\mathrm{d}\omega [\arg T(\omega)]$. The group delay dispersion was calculated with a least squares linear fit to the group delay profile.

The phase shifter settings were trained by minimizing the mean squared error between the realized and desired group delay profiles using the COBYLA optimization routine in \verb|SciPy| \cite{powell_direct_1998, scipy_10_contributors_scipy_2020}.

\begin{figure*}[bh]
    \centering
    \includegraphics[width=\textwidth]{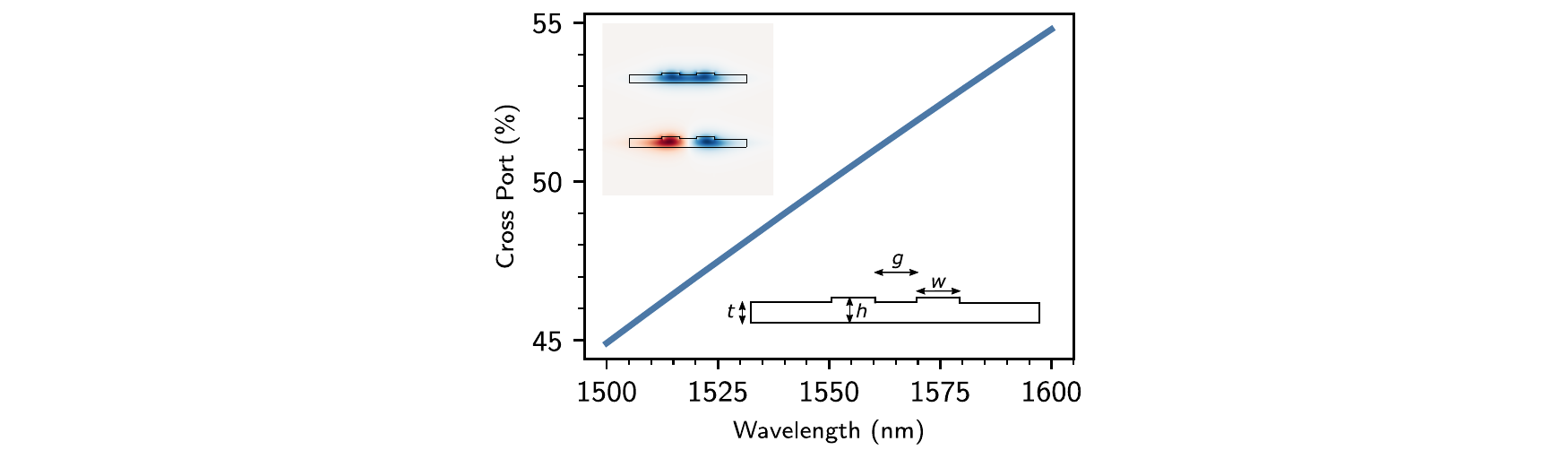}
    \caption{Wavelength vs.~cross coupling for the optimally tolerant directional coupler design ($w=400$ nm; $g = 400$ nm; $t = 150$ nm; $h = 220$ nm) in \cite{mikkelsen_dimensional_2014}.}
    \label{wl_dc}
\end{figure*}

\textbf{Unitary circuit bandwidth}: The wavelength dependence of the directional coupler design used in Figure 6b of the main text was computed with the MIT Photonic Bands (MPB) package \cite{johnson_block-iterative_2001}. Using MPB, we calculated the effective index difference $\Delta n(\lambda)$ between the first even and odd supermodes as a function of wavelength (Fig.~\ref{wl_dc}). Assuming the directional coupler is designed to operate as a 50-50 splitter at $\lambda_0$, the wavelength-dependent cross coupling is \cite{chrostowski_hochberg_2015}:
\begin{equation}
    T(\lambda) = \sin^2 \left [ \frac{\pi}{4} \left ( \frac{\Delta n(\lambda)}{\Delta n(\lambda_0)} \right ) \left (\frac{\lambda_0}{\lambda} \right)\right ]
\end{equation}
Figure 6b assumes $\lambda_0 = 1550$ nm.

\bibliography{supplement}